\documentclass[aps,pra,reprint, amsmath, amssymb,superscriptaddress,nofootinbib, nobibnotes, longbibliography]{revtex4-1}

\usepackage{bm}
\usepackage[retainorgcmds]{IEEEtrantools}
\usepackage{graphicx}
\usepackage{mathrsfs}
\usepackage{amsmath}
\usepackage{amsfonts}
\usepackage{amssymb}
\usepackage{color}
\usepackage{times,txfonts}
\usepackage{nicefrac}
\usepackage{ragged2e}
\usepackage[colorlinks=true,linkcolor=blue,urlcolor=blue,citecolor=blue,pdfusetitle]{hyperref}
\usepackage{braket}
\usepackage{enumitem}
\usepackage{lipsum}
\usepackage{dsfont} 
\usepackage[normalem]{ulem} 

\renewcommand{\Re}{\mathfrak{Re}}
\renewcommand{\Im}{\mathfrak{Im}}

\newcommand{\xaij}{\chi_{\alpha,ij}}
\newcommand{\Jaij}{J_{\alpha,ij}}
\newcommand{\xeff}{\xaij^{\rm eff}}
\newcommand{\Jeff}{J^{\rm eff}_{\alpha,ij}}
\newcommand{\Jxff}{J^{\rm eff}_{ij}}
\newcommand{\braemp}{\bra{\mathbf{0}}}
\newcommand{\ketemp}{\ket{\mathbf{0}}}
\newcommand{\ketone}[1]{\ket{\mathbf{1}_{#1}}}

\let\intop\relax
\let\iintop\relax
\let\iiintop\relax
\let\iiiintop\relax
\let\idotsintop\relax
\DeclareSymbolFont{cmex}{OMX}{cmex}{m}{n}
\DeclareMathSymbol{\intop}{\mathop}{cmex}{82}
\DeclareMathSymbol{\iintop}{\mathop}{cmex}{113}
\DeclareMathSymbol{\iiintop}{\mathop}{cmex}{114}
\DeclareMathSymbol{\iiiintop}{\mathop}{cmex}{115}
\DeclareMathSymbol{\idotsintop}{\mathop}{cmex}{116}
\let\int\intop

\newcommand{\expval}[1]{\left\langle #1 \right\rangle}
\DeclareMathOperator{\Tr}{Tr}

\newlength{\eqboxstorage}

\renewcommand\bra[1]{{\langle{#1}|}}
\makeatletter
\renewcommand\ket[1]{%
  \@ifnextchar\bra{\k@t{#1}\!}{\k@t{#1}}%
}
\newcommand\k@t[1]{{|{#1}\rangle}}

\newcommand{\editt}[1]{#1}
\newcommand{\edit}[1]{#1}
\newcommand{\stkout}[1]{}

\makeatother

\begin{document}

\title{Subtleties in the pseudomodes formalism}
\date{\today}
\author{Wynter Alford}
\email{Contact author: walford@ur.rochester.edu}
\affiliation{Department of Physics and Astronomy, University of Rochester, Rochester, New York 14627, USA}
\affiliation{University of Rochester Center for Coherence and Quantum Science, Rochester, New York 14627, USA}
\author{Laetitia P. Bettmann}
\affiliation{School of Physics, Trinity College Dublin, College Green, Dublin 2, D02K8N4, Ireland}
\author{Gabriel T. Landi}
\email{Contact author: gabriel.landi@rochester.edu}
\affiliation{Department of Physics and Astronomy, University of Rochester, Rochester, New York 14627, USA}
\affiliation{University of Rochester Center for Coherence and Quantum Science, Rochester, New York 14627, USA}
\begin{abstract}
The pseudomode method for open quantum systems, also known as the mesoscopic leads approach, consists in replacing a structured environment by a set of auxiliary ``pseudomodes'' subject to local damping that approximate the environment's spectral density.
Determining what parameters and geometry to use for the auxiliary modes, however, is non-trivial and involves many subtleties.
In this paper we revisit this problem of pseudomode design and investigate some of these subtleties.
In particular, we examine the scenario in which pseudomodes couple to each other, resulting in an effective spectral density that is no longer a sum of Lorentzians.
We show that non-diagonalizability of the pseudomodes’ effective single-particle non-Hermitian Hamiltonian can lead to terms in the effective spectral density which cannot be obtained by diagonalizable non-Hermitian Hamiltonians.
We also present a method for constructing the pseudomode parameters to exactly match a fit to a spectral density, and in doing so illuminate the enormous freedom in this process.
The case of many uncoupled pseudomodes evenly distributed in energy is explored, and we show how, contrary to conventional assumption, the effective spectral density does not necessarily converge in the limit of an infinite number of pseudomodes distributed this way\edit{.\stkout{; we attribute this to the non-completeness of Lorentzians as basis functions.}}
Finally, we discuss how the notion of effective spectral densities can also emerge in the context of scattering theory for non-interacting systems. 
\end{abstract}

\maketitle{}

\section{Introduction\label{sec:intro}}
No quantum system can ever be perfectly isolated from its environment, and so understanding the dynamics of open quantum systems is crucial across many fields. 
Many such systems are coupled sufficiently weakly to their environments that those environments can be approximated by simple Markovian Gorini-Kossakowski-Sudarshan-Lindblad (GKSL) master equations~\cite{gorini1976a, lindblad1976a}. 
However, many systems of interest couple strongly to and can build up quantum coherences with their environments, from quantum thermal machines~\cite{gelbwaser-klimovsky2015,thomas2018,camati2020,ptaszynski2022,latune2023} and
solid state devices~\cite{vanderwiel2000,fritz2005,kim2022,ye2024}
to light-matter systems both artificial~\cite{anappara2009,niemczyk2010,agarwal2013,forn-diaz2019,qin2024,das2025}
and biological~\cite{ishizaki2012,romero2014,chen2015,scholes2017}.
In all these cases, the evolution of the reduced system density matrix becomes non-Markovian, and more advanced techniques are required~\cite{tupkary2022}.

The method of pseudomodes~\cite{dalton2001,mazzola2009,subotnik2009,tamascelli2018, chen2019a,mascherpa2020a}, also known as mesoscopic leads~\cite{brenes2020a}, is one approach for treating the non-Markovian evolution of systems with strong system-bath coupling and/or highly structured spectral densities. 
In this method, the environment is replaced by a set of auxiliary modes subject to local damping. 
It was first shown that environments with a Lorentzian spectral density could be modeled by a single auxiliary mode with local damping by Imamo$\bar{\rm g}$lu~\cite{imamog-lu1994a, stenius1996a} and Garraway~\cite{garraway1997b, garraway1997c}, leading to the development of the reaction coordinate mapping technique~\cite{nazir2018, hughes2009} which has been applied extensively in quantum thermodynamics and related fields~\cite{iles-smith2014,strasberg2016,newman2017,strasberg2018,correa2019a, anto-sztrikacs2021, shubrook2025, mahadeviya2025a}. 
In the pseudomodes method, several auxiliary modes can be used with arbitrary couplings between them in order to represent an environment with an arbitrary spectral density. 
The original system and pseudomodes together form an ``extended system'' whose dynamics can then be modeled using a GKSL master equation. 
The dynamics of the original system's reduced density matrix will still be non-Markovian, while the dynamics of the full extended system remain Markovian; the pseudomode method can thus be thought of as a Markovian embedding~\cite{woods2014}.
This allows for quantum coherences to build up between the system and the bath, represented by coherences between the system and pseudomodes, without requiring the entire bath to be treated quantum mechanically. 

The method has been used to treat problems in areas including fermionic transport~\cite{chen2014a,schwarz2016,elenewski2017a,rams2020a,wojtowicz2021, elenewski2021,brenes2023a, cirio2023}, quantum thermodynamics~\cite{lacerda2023b,lacerda2024,bettmann2025,menczel2024a,albarelli2024}, quantum optics~\cite{feist2020,medina2021,sanchez-barquilla2022,lednev2024a, liang2024}\edit{, and strongly-correlated systems~\cite{arrigoni2013,dorda2014,dorda2015}.}
Recent works have combined this method with tensor networks~\cite{brenes2020a,cirio2023, lacerda2024} and the Time Evolving Density operator with Orthogonal Polynomials Algorithm (TEDOPA)~\cite{nusseler2022, ferracin2024b} to treat interacting systems. 
One advantage of this method is that by modeling the environments using physical modes to form an extended system, we have access to details about correlations between the system and its environment at different energies. 
This has been used to gain insight into energy-resolved currents for interacting fermionic systems~\cite{brenes2020a}. 
Likewise, using the framework of continuous measurement, one can monitor quantum jumps between the residual baths and pseudomodes and perform trajectory unraveling to probe stochastic thermodynamic quantities~\cite{bettmann2025}. 
The auxiliary modes can therefore possess physical significance, rather than serving merely as a mathematical construct for modeling the environment.
Alternatively, one can relax the constraint that the pseudomodes have physical parameters and include non-Hermitian terms in the pseudomode Hamiltonian or negative damping rates in the GKSL master equation\edit{~\cite{lambert2019,luo2023,menczel2024a,pleasance2020a}}, or add additional dissipators acting non-locally on the system and pseudomodes~\cite{park2024,huang2025}, in order to model an even wider variety of environments.

The success of the pseudomodes approach, however, relies on the careful design of their geometry and parameters so as to properly mimic the real environment. 
Implementing this entails various subtleties, which are often overlooked. 
The goal of this paper is to address said subtleties.
In section~\ref{sec:computeJeff}, we consider the effective spectral density that arises from a particular configuration of pseudomodes. Depending on the non-Hermitian Hamiltonian generated by the pseudomodes' Hamiltonian and the residual baths, three possible cases emerge. 
If none of the pseudomodes couple to each other, then one obtains an effective spectral density that is a sum of Lorentzians, while if the pseudomodes couple to each other there are additional terms in the effective spectral density which we call ``Anti-Lorentzian''.
We take this one step further, and show that one can obtain terms which are neither Lorentzian nor Anti-Lorentzian by using a non-diagonalizable non-Hermitian Hamiltonian for the pseudomodes\edit{, a case which to the best of our knowledge has not been previously \editt{characterized in detail.}}

In section~\ref{sec:fitting} we discuss the reverse process of choosing pseudomode parameters to obtain a particular effective spectral density, which is required to actually implement the pseudomode method. 
This is generally done in two steps. 
First, the true spectral density is fit in \edit{time} space with a set of decaying complex exponentials.
Then, the pseudomode parameters must be chosen to match that exponential fit, which is typically done by imposing some additional structure on the pseudomode parameters and then performing optimization~\cite{mascherpa2020a}.
This second step is the greatest challenge to implementing the pseudomode method, in large part because there are many more free parameters for the pseudomodes than there are constraints given by the effective spectral density. 
We present a new method for choosing the pseudomode parameters, by which the parameters can be chosen to exactly match a decaying exponential fit.
Our method also sheds light on the vast number of free parameters involved when choosing the pseudomode parameters.

In section~\ref{sec:mesofitting}, we briefly explore the use of uncoupled pseudomodes uniformly distributed in energy and show that, contrary to conventional assumption, the effective spectral density obtained from this approach approximates but does not converge to the true bath spectral density. 
Instead, we present a new, alternate method which , while also not convergent in the infinite limit, significantly reduces the error between the true and effective spectral densities in the limit of many pseudomodes and is equally easy to implement. 
Finally, in section~\ref{sec:scattering} we show how the effective spectral densities of section~\ref{sec:computeJeff} also emerge in the context of scattering theory for non-interacting systems~\cite{blanter2000a, buttiker1992a, levitov1996a, esposito2009b}. 
We confirm that if the pseudomode parameters are chosen such that the effective spectral density exactly matches the true spectral density then the currents obtained from Landauer-Büttiker theory will also match exactly between the original system-bath description and pseudomode description.
This also presents a new way to study scattering theory currents by breaking down a current between two real baths into the sum of currents between residual baths under a pseudomode mapping.

\subsection{Model System\label{ssec:model}}
\begin{figure*}[!th]
    \centering
    \includegraphics[width=\linewidth]{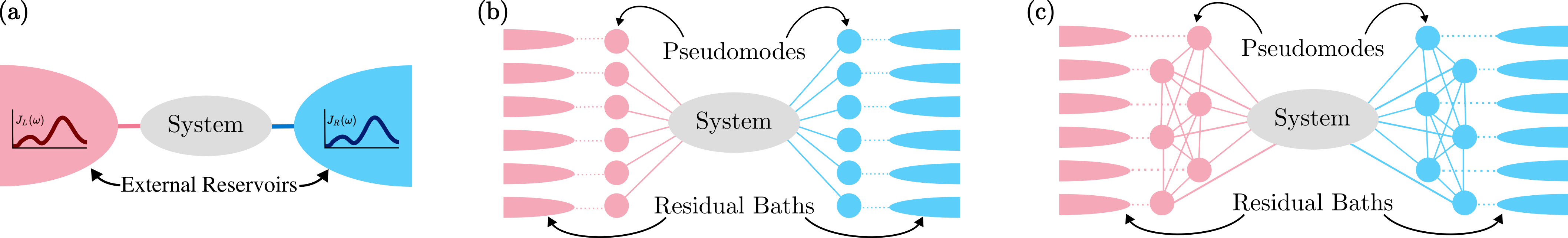}
    \caption{(a) Standard setup of a quantum system coupled to two reservoirs. (b) The pseudomodes model with uncoupled pseudomodes (the Diagonal case), where each bath has been replaced by a finite set of damped pseudomodes. (c) The general pseudomode model, where the pseudomodes are allowed to couple to each other arbitrarily in addition to the residual baths.}
    \label{fig:drawing}
\end{figure*}
We consider the system depicted in Fig.~\ref{fig:drawing}(a), consisting of a fermionic system coupled to multiple reservoirs. 
For simplicity, we take two reservoirs $L,R$ (left and right), but the result generalizes for more than two.
The system is described by $n_S$ fermionic modes $c_i$. The total Hamiltonian of the system and environment is:
\begin{equation}\label{eqn:H_basic}
\begin{aligned}
    H_{\rm tot} =& H_S(t) + \sum_{\alpha \in \{L,R\}} \sum_{k\edit{\in\alpha}}  \omega_k d_{\alpha k}^\dagger d_{\alpha k}^{}  
    \\[0.2cm]
    &+ \sum_{\alpha \in \{L,R\}} \edit{\sum_{i\in S}\sum_{k\in\alpha}}\left(  g_{\alpha i,k} c_i^\dagger d_{\alpha k}^{} + \edit{g_{\alpha i,k}^*} d_{\alpha k}^\dagger c_i^{}\right)\,,
\end{aligned}    
\end{equation}
where $d_{\alpha k}$ is the $k$th fermionic mode of bath $\alpha \in \{L,R\}$.
We set $\hbar=1$ here and throughout the paper.
The system Hamiltonian can be completely general and may include arbitrary driving. 
The last term describes how bath $\alpha \in \{L,R\}$ can, in principle, couple to any mode $c_i$ of the system.
Each bath is characterized by a spectral density matrix 
\begin{equation}\label{eqn:J_basic}
    (J_\alpha)_{ij} = 2\pi \sum_{k\edit{\in\alpha}} g_{\alpha i,k}^{\edit{*}} g_{\alpha j,k}^{} \delta(\omega-\omega_k)\,,
\end{equation}
which is a real and symmetric matrix of dimension $n_S\times n_S$.

The equation of motion for $c_i(t)$, in the Heisenberg picture, reads
\begin{equation}\begin{aligned}
    \frac{dc_i(t)}{dt}=i[H_S(t),c_i(t)]-i\sum_{\alpha}\edit{\sum_{k\in\alpha}}g_{\alpha i,k}e^{-i\omega_{\alpha k}t}d_{\alpha k}\\
    -\sum_{\alpha}\edit{\sum_{k\in\alpha}\sum_{j\in S}}\int_0^tdt'g_{\alpha i,k}^{\edit{*}}g_{\alpha j,k}^{}e^{-i\omega_{\alpha k}(t-t')}c_j(t')\, ,
\end{aligned}\end{equation}
which we can express as a quantum Langevin equation, in terms of noise operators $\xi_{\alpha i}$ and memory kernel matrices $\xaij$, as~\cite{brenes2020a}:
\begin{equation}\begin{aligned}\label{eqn:langevin_basic}
    \frac{dc_i(t)}{dt}&=i[H_S(t),c_i(t)]+\sum_\alpha\xi_{\alpha i}(t)\\&-\sum_{\alpha}\edit{\sum_{j\in S}} \int_0^tdt'\chi_{\alpha,ij}(t-t')c_j(t')
\end{aligned}\end{equation}
where $\xi_{\alpha i}=-i\sum_{k\edit{\in\alpha}}g_{\alpha i,k}e^{-i\omega_{\alpha k}t}d_{\alpha k}$ is the noise operator, and the memory kernel is given by the Fourier transform of the spectral density:
\begin{equation}\label{eqn:XasFT}
    \chi_{\alpha,ij}(t)=\int\frac{d\omega}{2\pi}J_{\alpha,ij}(\omega)e^{-i\omega t}\,.
\end{equation}

While only the $t>0$ part of the memory kernel enters into the dynamics governed by~\eqref{eqn:langevin_basic}, we define the memory kernel for both positive and negative time using~\eqref{eqn:XasFT} so that we can easily convert back and forth between $\Jaij$ and $\xaij$.

The open system dynamics described by~\eqref{eqn:langevin_basic} are fully determined by the operators $\xi_{\alpha i}(t)$ and $\chi_{\alpha,ij}(t)$. Alternately, one can think of the dynamics as determined by the bath correlation functions:
\begin{align}\label{eqn:correlations_basic}
    &C^+_{\alpha,ij}(t)=\int\frac{d\omega}{2\pi}e^{i\omega t}J_{\alpha,ij}(\omega)f_\alpha(\omega)\,,\\
    &C^-_{\alpha,ij}(t)=\int\frac{d\omega}{2\pi}e^{-i\omega t}J_{\alpha,ij}(\omega)\big(1-f_\alpha(\omega)\big)\, ,
\end{align}
where $f_\alpha(\omega)$ is the Fermi function associated to bath $\alpha$. From these two expressions we can then see that the dynamics can also be thought of as fully determined by the spectral density matrices $\Jaij(\omega)$~\eqref{eqn:J_basic} and the Fermi functions $f_\alpha(\omega)$ of the baths. This is the most useful description for the implementation of the pseudomode method.

Finally, it is worth noting that if we express the system-bath interaction Hamiltonian as:
\begin{equation}
    \sum_\alpha\sum_{i\edit{\in S}}c_i^\dag B_{\alpha i}^{}+B_{\alpha i}^\dag c_i^{}\,,
\end{equation}
with $B_{\alpha i}=\sum_kg_{\alpha i,k}d_{\alpha k}$, then the two correlation functions can equivalently be written as:
\begin{align}\label{eqn:C+expectation}
    &C^+_{\alpha\edit{,} ij}(t)=\expval{B_i^\dag(t)B_j^{}(0)}\,,\\
    &C^-_{\alpha\edit{,} ij}(t)=\expval{B_i^{}(t)B_j^\dag(0)}\label{eqn:C-expectation}
    \,.
\end{align}

\subsection{Pseudomode Mapping\label{ssec:mesomodel}}
Under the pseudomode method, each bath $\alpha\in\{L,R\}$ is replaced by a finite number $n_\alpha$ of fermionic pseudomodes $a_{\alpha k}$ (sometimes called lead modes or auxiliary modes), each of which is coupled to its own residual bath, as depicted in Fig.~\ref{fig:drawing}(c).
The Hamiltonian in~\eqref{eqn:H_basic} is mapped to:

\begin{align}\label{eqn:H_Meso}
    &H_{\rm tot} = H_S(t) + 
    \sum_{\alpha \in \{L,R\}} \sum_{k,q\edit{=1}}^{\edit{n_\alpha}}  \Lambda_{\alpha, kq} a_{\alpha k}^\dagger a_{\alpha q}^{}\nonumber
    \\[0.2cm]
    &\qquad + \sum_{\alpha \in \{L,R\}} \edit{\sum_{i\in S}\sum_{k=1}^{n_\alpha}} \big(\zeta_{\alpha i,k} a_{\alpha k}^\dagger c_i^{} + \zeta_{\alpha i,k}^* c_i^\dagger a_{\alpha k}^{}\big)
    \\[0.2cm]
    &+ \sum_{\alpha \in \{L,R\}} \sum_{k\edit{=1}}^{\edit{n_\alpha}} \sum_{p\edit{\in(\alpha,k)}}\Big\{ 
    \omega_p b_{\alpha k, p}^\dagger b_{\alpha k, p}^{} + \tau_{\alpha k, p}\big( a_{\alpha k}^\dagger b_{\alpha k, p}^{} + \edit{b_{\alpha k, p}^{\dag}} a_{\alpha k}^{}\big)\Big\}\,.\nonumber
\end{align}

The first line of~\eqref{eqn:H_Meso} describes the possible interactions between pseudomodes of the same bath, described by the $n_\alpha\times n_\alpha$ matrix $\Lambda_\alpha$ for each bath. 
If $\Lambda$ is diagonal then none of the pseudomodes couple to each other, and we have the simpler configuration shown in Fig.~\ref{fig:drawing}(b).
The second line describes how each pseudomode interacts with the central system through a set of $n_\alpha \times n_S$ matrices $\zeta_\alpha$. 
Finally, the last line describes a set of infinite-dimensional residual reservoirs, one for each pseudomode $a_{\alpha k}$, \edit{labeled by the pair of indices $(\alpha,k)$ and} composed of a set of modes $b_{\alpha k, p}$. 
The residual baths then have spectral densities given by
\begin{equation}
    J_{\alpha k}^{\rm res}(\omega)=2\pi\sum_{p\edit{\in(\alpha,k)}} \left|\tau_{\alpha k, p}\right|^2\delta(\omega-\omega_p)\, ,
\end{equation}
which is a scalar, rather than a matrix like~\eqref{eqn:J_basic}, since each residual bath $(\alpha,k)$ couples only to the pseudomode $a_{\alpha k}$.
Generally, we want the couplings to the residual baths $\tau_{\alpha k, p}$ to be constant at all energies, so that the residual baths have constant spectral densities 
\begin{equation}\label{eqn:flat-residuals}
    J^{\rm res}_{\alpha k}(\omega)=\Gamma_{\alpha\edit{,kk}}\,.
\end{equation}
\edit{This allows us to treat the ``extended system'' consisting of the original system and the pseudomodes using a Markovian master equation.} In this case, the \edit{(original)} system feels an effective memory kernel given by~\cite{mascherpa2020a}:
\begin{equation}\label{eqn:zeWtz}
    \xeff(t)=\begin{cases}
        \zeta_i^\dag e^{Wt}\zeta_j&t\ge0\\
        \zeta_j^\dag e^{-W^\dag t}\zeta_i&t\leq0\,.
    \end{cases}
\end{equation}
where
\begin{equation}\label{eqn:W_matrix}
    W_\alpha=-i\Lambda_\alpha-\Gamma_\alpha/2\,.
\end{equation}
Once again, only the $t\geq 0$ part of $\xeff$ enters into the dynamics, but we define the $t<0$ part by $\xaij(-t)=\xaij(t)^*$ based on ~\eqref{eqn:XasFT}, so that we can take the Fourier transform and conclude that the system feels an effective spectral density given by:
\begin{equation}\label{eqn:Jeff}
    J^{\rm eff}_{\alpha,ij}(\omega)=2\Im\left(\zeta^\dag_{\alpha i}\frac{1}{iW_\alpha-\omega}\zeta_{\alpha j}\right)\,.
\end{equation}

\edit{The derivation for these forms of $\Jeff,\xeff$ can be found in Appendix~\ref{apx:Jeffderivation}.} 
The pseudomode parameters are then chosen so that the effective spectral density~\eqref{eqn:Jeff} and effective memory kernel~\eqref{eqn:zeWtz} are good approximations for the true spectral density~\eqref{eqn:J_basic} and memory kernel~\eqref{eqn:XasFT} of the original system.
\edit{\stkout{The derivation for these forms of $\Jeff,\xeff$ can be found in Appendix~\ref{apx:Jeffderivation}.}} 
The reduced system dynamics under~\eqref{eqn:H_basic} will then be identical to those under~\eqref{eqn:H_Meso} so long as the correlation functions from $J^{\rm eff}_{\alpha, ij}$ are identical to the true correlation functions~\eqref{eqn:correlations_basic}; likewise, the dynamics under~\eqref{eqn:H_basic} will be well-approximated by the dynamics under~\eqref{eqn:H_Meso} so long as the effective correlation functions well-approximate the true ones~\cite{trivedi2021a,huang2024}. 
In practice, we achieve this by ensuring that $\Jeff(\omega)$ is a good fit to $\Jaij(\omega)$\edit{, while setting the temperatures and chemical potentials of the residual baths $(\alpha,k)$ to be the same as those of the original baths $\alpha$.}

The utility of this method thus hinges on our ability to choose a set of $\zeta_{\alpha i}$, $\Lambda_{\alpha}$, and $\Gamma_{\alpha}$ so that the effective spectral density~\eqref{eqn:Jeff} closely matches the actual spectral density~\eqref{eqn:J_basic}. 
In the next section, we describe how to obtain the form of $\Jeff$ from the chosen pseudomode configuration; in the following section, we describe the more complicated problem of choosing pseudomode parameters to obtain a desired $\Jeff$.
As we shall show shortly, there are many possible choices of pseudomode parameters that give rise to the same effective spectral density, and not all of them may be equally desirable. 
Actually computing the extended-system dynamics generally requires further approximations, most commonly the use of a local GKSL master equation, which may be more or less reasonable to make based on the choice of pseudomode parameters. 
We do not discuss the merits of particular choices of parameters in this paper, instead focusing on the procedure for constructing the pseudomode parameters and emphasizing the freedoms therein. 
\edit{Note also that while in this paper we focus on fermionic systems, the expressions~\eqref{eqn:zeWtz} and~\eqref{eqn:Jeff} are identical for bosonic systems under the rotating wave approximation, where the couplings between the system and residual bath take the form~\eqref{eqn:H_basic}~\cite{mascherpa2020a,medina2021}, as are the results of section~\ref{sec:computeJeff}, which follow from them. 
For bosonic systems with position-position coupling (i.e, $(c+c^\dag)(d+d^\dag)$), one obtains slightly different results, but the general ideas of this paper may remain applicable.
Some additional care is required when treating bosonic systems, however, since~\eqref{eqn:Jeff} will typically produce an effective spectral density which is nonzero for $\omega<0$.}

\section{Computation of Effective Spectral Density\label{sec:computeJeff}}
The form that the effective spectral density~\eqref{eqn:Jeff} of bath $\alpha$ represented by \edit{$n_\alpha$} pseudomodes takes depends on whether the matrix $W_\alpha$~\eqref{eqn:W_matrix} is diagonal, diagonalizable, or non-diagonalizable. 
A pseudomode configuration is said to be diagonal, diagonalizable or non-diagonalizable when its matrix $W_\alpha$ has the corresponding property. 
These cases are described sequentially below. 
In general we might also have a matrix $W_\alpha$ which is block-diagonal, where some blocks are diagonalizable and some are not; in this case, each block can be treated separately and the resulting effective spectral density is merely the sum of the effective spectral densities from each block.

\edit{While all the results shown in this section follow straightforwardly from~\eqref{eqn:zeWtz} and~\eqref{eqn:Jeff}, we show them explicitly, as we will use these results for the remainder of the paper. Writing these results explicitly also makes clear the difference in the effective spectral densities that can be obtained using coupled modes as opposed to uncoupled modes, and why it is often desirable to use coupled modes. Additionally, \editt{we explore the appearance of new terms in the effective spectral density when the pseudomodes' non-Hermitian Hamiltonian is non-diagonalizable.}} 

\subsection{Diagonal Case\label{ssec:diagonal}}

If the pseudomode-pseudomode interaction matrix $\Lambda_\alpha$ is diagonal, this means that each pseudomode corresponding to bath $\alpha$ couples to the system but not to any other pseudomodes, as depicted in Fig.~\ref{fig:drawing}(b). 
In this case $e^{W_\alpha t}$ is also diagonal. 
If we label the on-site energies of each pseudomode as $\Lambda_{\alpha,jj}=\varepsilon_{\alpha j}$, and the coupling of each one to their residual baths as $\Gamma_{\alpha,jj}=\gamma_{\alpha j}$, then the resulting effective memory kernel~\eqref{eqn:zeWtz} is a sum of complex exponentials:
\begin{equation}\label{eqn:Xeff_diagonal}
    \xeff(t)=\begin{cases}
        \edit{\sum_{k=1}^{n_\alpha}}\zeta_{\alpha i,k}^*\zeta_{\alpha j,k}^{}e^{-i\varepsilon_{\alpha k}t-\gamma_{\alpha k}t/2} & t\geq0\\
        \edit{\sum_{k=1}^{n_\alpha}}\zeta_{\alpha i,k}^{}\zeta_{\alpha j,k}^*e^{-i\varepsilon_{\alpha k}t+\gamma_{\alpha k}t/2} & t\leq0\,.
    \end{cases}
\end{equation}

Taking the Fourier transform, the resulting effective spectral density matrix~\eqref{eqn:Jeff} is then:
\begin{equation}\label{eqn:Jeff_diagonal}
    \Jeff(\omega)=\edit{\sum_{k=1}^{n_\alpha}} \frac{\Re\left(\zeta_{\alpha i,k}^*\zeta_{\alpha j,k}^{}\right)\gamma_{\alpha k}-2\Im\left(\zeta_{\alpha i,k}^*\zeta_{\alpha j,k}^{}\right)(\omega-\varepsilon_{\alpha k})}{(\omega-\varepsilon_{\alpha k})^2+(\gamma_{\alpha k}/2)^2}\,.
\end{equation}

In particular the diagonal elements are:
\begin{equation}\label{eqn:Jeff_diagonal_simple}
    J^{\rm eff}_{\alpha,ii}(\omega)=\edit{\sum_{k=1}^{n_\alpha}} \frac{\left|\zeta_{\alpha i,k}\right|^2\gamma_{\alpha k}}{(\omega-\varepsilon_{\alpha k})^2+(\gamma_{\alpha k}/2)^2}\,,
\end{equation}
which is a sum of Lorentzians with each pseudomode contributing one Lorentzian, with the form:
\begin{equation}\label{eqn:lorentzian}
    \frac{\gamma_{\alpha k}}{(\omega-\varepsilon_{\alpha k})^2+(\gamma_{\alpha k}/2)^2}\,.
\end{equation}
If each bath couples to only one system site, then the effective spectral densities for each bath will contain only a sum of Lorentzians. 
Any off-diagonal elements contain both Lorentzian terms~\eqref{eqn:lorentzian} and terms of the form
\begin{equation}\label{eqn:antilorentzian}
    \frac{(\omega-\varepsilon_{\alpha k})}{(\omega-\varepsilon_{\alpha k})^2+(\gamma_{\alpha k}/2)^2}\,,
\end{equation}
which we call ``Anti-Lorentzian'' since they structurally resemble Lorentzians but are anti-symmetric about the energy $\varepsilon$, rather than peaked there.
\begin{figure*}[!t]
    \centering
    \includegraphics[width=0.33\linewidth]{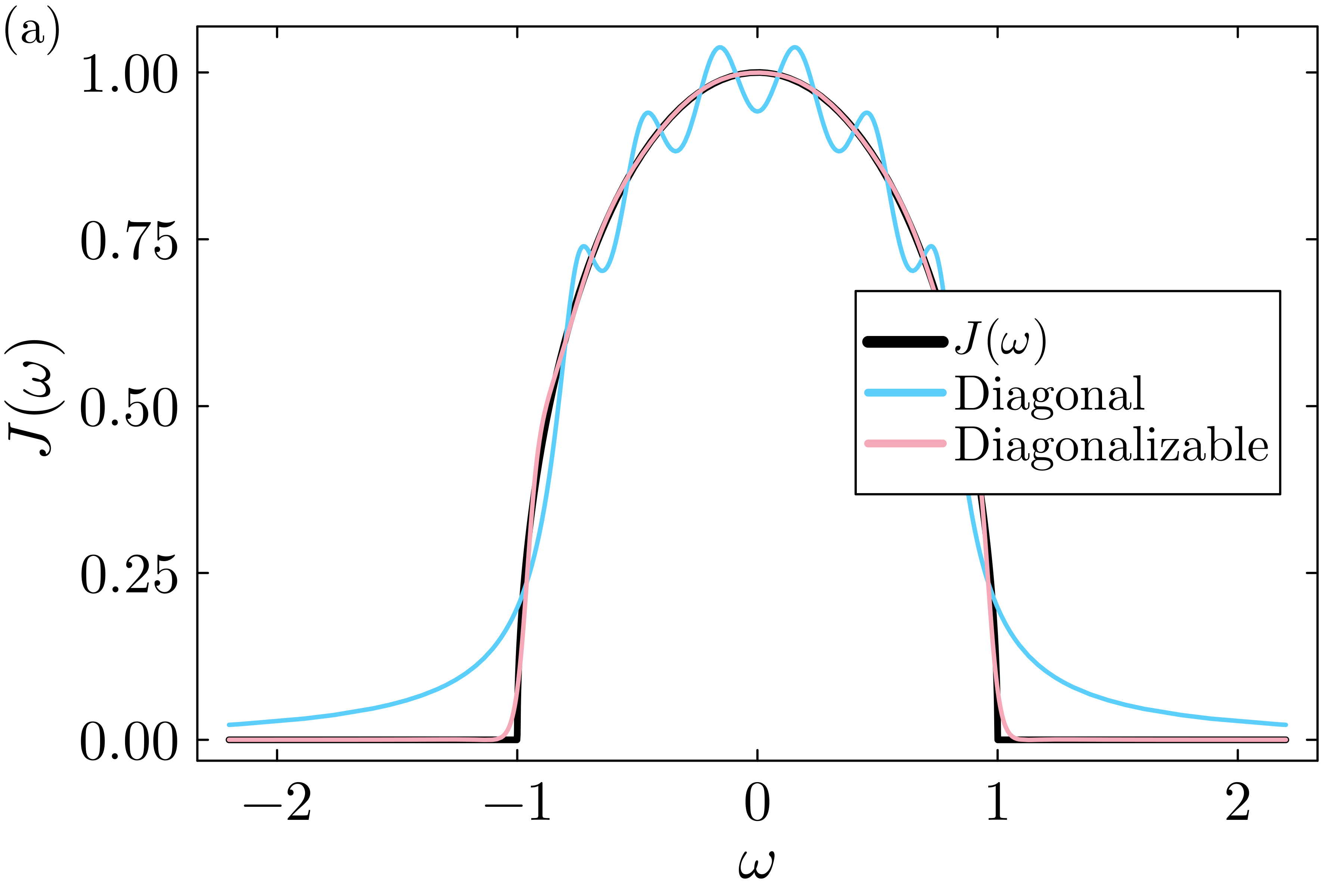}
    \includegraphics[width=0.33\linewidth]{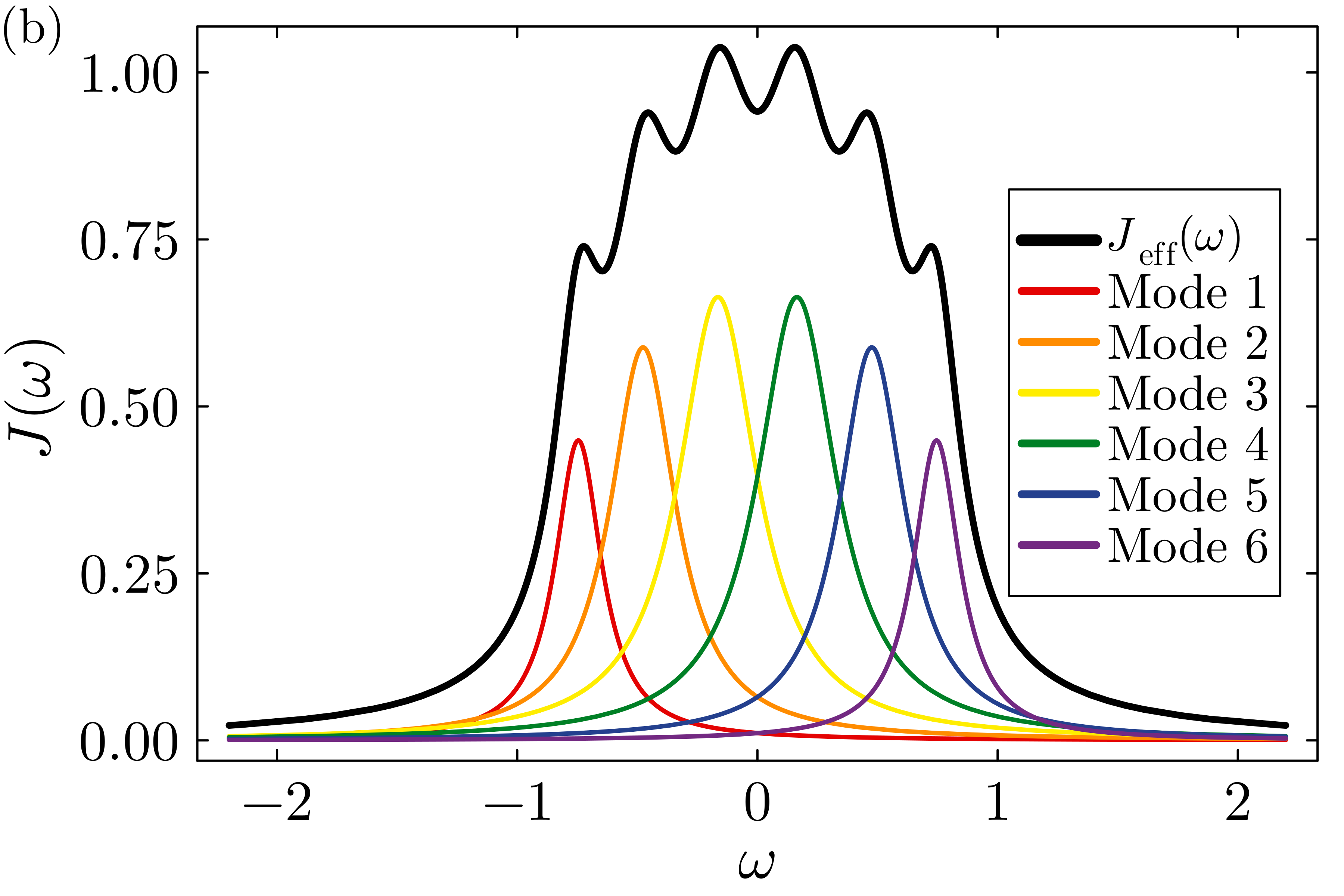}
    \includegraphics[width=0.33\linewidth]{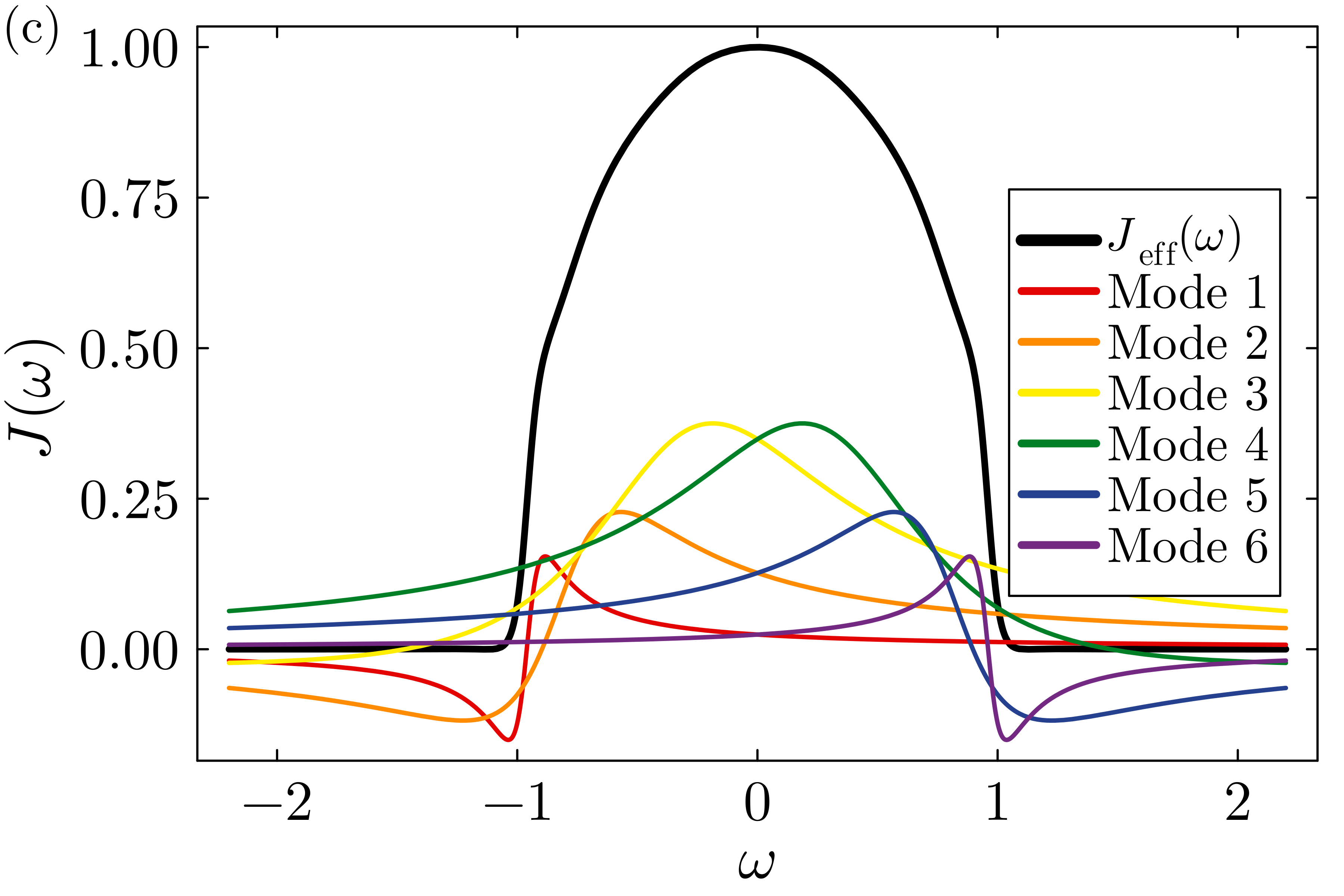}
    \caption{(a) Comparison of fits to an effective spectral density from a diagonal configuration (sum of Lorentzians, with uncoupled pseudomodes) and from a diagonalizable one (sum of Lorentzians and Anti-Lorentzians, with pseudomodes coupled to one another). The true spectral density is $J(\omega)=\sqrt{1-\omega^2}$, which both fits model using 6 pseudomodes. The diagonal-case fit was obtained by brute-force optimization for the parameters $|\zeta_k|^2,\varepsilon_k,\gamma_k$, while for the diagonalizable case, Prony's method (described in section~\ref{sec:fitting} and appendix~\ref{apx:prony}) was used to obtain the best-fit $\kappa_k,\varepsilon_k,\gamma_k$. (b) Contributions to the effective spectral density in the diagonal case from each of the 6 pseudomodes; each mode contributes a Lorentzian term. (c) Contributions to the effective spectral density in the diagonalizable case from each of the 6 pseudomodes; each mode contributes the sum of a Lorentzian term and an Anti-Lorentzian term. The parameters from optimization used in these fits are given in Appendix~\ref{apx:data}.}
    \label{fig:comparefits}
\end{figure*}

\subsection{Diagonalizable Case\label{ssec:diagonalizable}}
Suppose the matrix $W_\alpha$ is not diagonal but can be diagonalized as $W_\alpha=S_\alpha M_\alpha S_\alpha^{-1}$ with $M_\alpha$ a diagonal matrix. 
Note that $S_\alpha$ is not unitary since $W_\alpha$ is not Hermitian. 
The effective memory kernel matrix~\eqref{eqn:zeWtz} is then given by:
\begin{equation}
    \xeff(t)=\zeta_{\alpha i}^\dag S_\alpha e^{{M_\alpha}t}S_\alpha^{-1}\zeta_{\alpha j}
    \,.
\end{equation}
Labeling the eigenvalues of $W_\alpha$ as $-i\varepsilon_j-\gamma_j/2$ and writing out the multiplication explicitly, we obtain:
\begin{equation}\label{eqn:Xeff_diagonalizable}
    \xeff(t)=\begin{cases}
        \edit{\sum_{k=1}^{n_\alpha}}\kappa_{\alpha k,ij}e^{-i\varepsilon_{\alpha k}t-\gamma_{\alpha k}t/2} & t\geq0\\
        \edit{\sum_{k=1}^{n_\alpha}}\kappa_{\alpha k,ij}^*e^{-i\varepsilon_{\alpha k}t+\gamma_{\alpha k}t/2} & t\leq0
    \end{cases}\,,
\end{equation}
with the amplitudes
\begin{equation}\label{eqn:kappa}
    \kappa_{\alpha k,ij}=\sum_{p,q\edit{=1}}^{\edit{n_\alpha}}\zeta_{\alpha i,p}^*\zeta_{\alpha j,q}S_{\alpha, pk}S_{\alpha, kq}^{-1}\,.
\end{equation}
Note that the diagonal elements $\kappa_{\alpha k,ii}$ \edit{may be complex, but} always sum to a positive real number:
\begin{equation}\label{eqn:kappa_sum}
    \sum_{k\edit{=1}}^{\edit{n_\alpha}}\kappa_{\alpha k,ii}>0\,.
\end{equation}
\edit{This fact will become a key constraint when choosing pseudomode parameters to fit a spectral density.}

 Under this model, even if bath $\alpha$ couples to only one system site, it is still possible to obtain $\mathfrak{Im}(\kappa_{\alpha k,ii})\neq 0$. The effective spectral density~\eqref{eqn:Jeff} thus has a similar form to the diagonal case, but because the amplitudes~\eqref{eqn:kappa} are more complicated, it can now have both Lorentzian~\eqref{eqn:lorentzian} and Anti-Lorentzian~\eqref{eqn:antilorentzian} terms in both its diagonal and off-diagonal elements:
\begin{equation}\label{eqn:Jeff_diagonalizable}
    \Jeff(\omega)=\edit{\sum_{k=1}^{n_\alpha}} \frac{\Re\left(\kappa_{\alpha k,ij}\right)\gamma_{\alpha k}-2\Im\left(\kappa_{\alpha k,ij}\right)(\omega-\varepsilon_{\alpha k})}{(\omega-\varepsilon_{\alpha k})^2+(\gamma_{\alpha k}/2)^2}\,.
\end{equation}

Using a diagonalizable but not diagonal $W$ often allows for a much better fit to structured spectral densities than using a diagonal $W$, as the negativities and anti-symmetry from the Anti-Lorentzian terms allow one to capture features much more easily. 
Figure~\ref{fig:comparefits} compares 6-mode fits to the semi-elliptical spectral density, showing how the diagonalizable version matches the true spectral density much more closely, particularly around the sharp corners at $\omega=\pm 1$. \editt{For a general memory kernel, the fitting error $\epsilon$ over a time window $T$ scales as $\mathrm{poly}(T/\epsilon)$ for uncoupled (diagonal) modes, but $\mathrm{polylog}(T/\epsilon)$ for diagonalizably coupled modes~\cite{huang2025,thoenniss2024}.}

\subsection{Non-diagonalizable Case\label{ssec:non-diagonalizable}}

So long as $W_\alpha$ is diagonalizable, we can only end up with Lorentzian~\eqref{eqn:lorentzian} and Anti-Lorentzian~\eqref{eqn:antilorentzian} terms in the spectral density matrix. 
However, in the case that $W_\alpha$ is non-diagonalizable (sometimes called ``defective''), \edit{which in the language of non-Hermitian Hamiltonians corresponds to an exceptional point in the pseudomodes' non-Hermitian Hamiltonian $iW$}, other functional forms can appear.

An $n\times n$ matrix is non-diagonalizable if it has fewer than $n$ unique eigenvectors, and is ``maximally defective'' if it has only one eigenvector.
If $W_\alpha$ is non-diagonalizable, it can be written in Block Jordan form. 
Each eigenvalue of $W_\alpha$ corresponds to a $d_i\times d_i$ block, where $d_i$ is the algebraic multiplicity of that eigenvalue. 
For simplicity suppose that $W_\alpha$ has one maximally defective eigenvalue $w$; the general result then follows. 
In that case we can write

\begin{equation}
    W=S(w\mathbf{1}+N)S^{-1}=w\mathbf{1}+SNS^{-1}\,,
\end{equation}
where $N_{ij}=\delta_{i,j-1}$ is a matrix with all $1$s on the superdiagonal and zeros elsewhere. 
Then $e^{Wt}=e^{wt}Se^{Nt}S^{-1}$ where the Taylor series for $e^N$ only contains $n$ terms since $N$ is nilpotent. 
We can use this to compute the form of $e^N$ as:
\begin{equation}
\left(e^{Nt}\right)_{i\geq j}=\frac{t^{i-j}}{(i-j)!}\,.
\end{equation}
The resulting contribution to the effective memory kernel~\eqref{eqn:zeWtz} (for $t\geq0$) is then:
\begin{equation}\label{eqn:Xeff_nondiagonalizable}
    \xeff(t)=e^{wt} \sum_{p,q\edit{=1}}^{\edit{n_\alpha}}\sum_{k=1}^{\edit{n_\alpha}}\sum_{l=1}^k\frac{1}{(k-l)!} \zeta_{\alpha i,p}^*\zeta_{\alpha j,q} S_{pk} S^{-1}_{lq}t^{k-l} \,.
\end{equation}

We can see in particular that since $1\leq l\leq k\leq d$, the highest possible power of $t$ that can appear inside the summation is $t^{d-1}$. 
The general form of the memory kernel for $t\geq0$ is thus
\begin{equation}
    \chi(t)\sim\sum_{k=0}^{d-1} \kappa_k t^k e^{wt}\,.
\end{equation}
The terms in the memory kernel with $t^k\exp(...)$ then give rise to terms in the spectral density with terms like \begin{equation}
    \frac{1}{(\omega^2+\gamma^2)^{k+1}}\,.
\end{equation}
Examples of the types of terms that arise are shown in Table~\ref{tab:ndterms}. Note that while there are known methods for constructing ${n\times n}$ non-diagonalizable matrices~\cite{scott1997}, the form of~\eqref{eqn:Xeff_nondiagonalizable} makes it difficult to construct such matrices that will give rise to a memory kernel that is an arbitrary linear combination of such terms.

\begin{table}[!b]
\caption{Contribution to the effective spectral density from terms $t^{k}e^{wt}$ in the effective memory kernel for $t>0$. 
Note that terms with real and imaginary coefficients give different functional forms since ${\chi(-t)=-\chi(t)^*}$.}
\label{tab:ndterms}
\begin{ruledtabular}
\begin{tabular}{@{}c@{}c@{}}
Term in $\chi_{\rm eff}(t)$ & Term in $J_{\rm eff}(\omega)$ \\
\colrule
$e^{-\gamma t}$ & $\frac{2\gamma}{\omega^2+\gamma^2}$ \\\hline
$-ie^{-\gamma t}$ & $\frac{2\omega}{\omega^2+\gamma^2}$ \\\hline
$te^{-\gamma t}$ & $\frac{2(\gamma^2-\omega^2)}{(\omega^2+\gamma^2)^2}$ \\\hline
$-ite^{-\gamma t}$ & $\frac{4\gamma\omega}{(\omega^2+\gamma^2)^2}$ \\\hline
$t^2e^{-\gamma t}$ & $\frac{4(\gamma^3-3\gamma\omega^2)}{(\omega^2+\gamma^2)^3}$ \\\hline
$-it^2e^{-\gamma t}$ & $\frac{4(3\gamma^2\omega-\omega^3)}{(\omega^2+\gamma^2)^3}$ \\\hline
$t^n e^{-\gamma t}$ & $n! \left[(\gamma +i \omega )^{-n-1}+(\gamma -i \omega )^{-n-1}\right]$ \\\hline
$-it^n e^{-\gamma t}$ & $-i n! \left[(\gamma -i \omega )^{-n-1}-(\gamma +i \omega )^{-n-1}\right]$ \\
\end{tabular}
\end{ruledtabular}
\end{table}

\subsubsection*{Example: 2$\times$2 non-diagonalizable Case\label{sssec:ND2}}
\begin{figure}[!tb]
    \centering
    \includegraphics[width=0.7\linewidth]{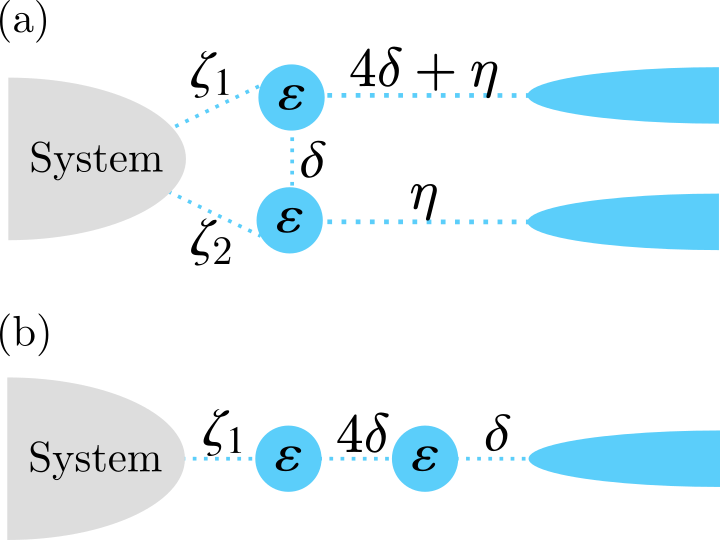}
    \caption{\edit{(a) The most general non-diagonalizable pseudomode configuration, consisting of two pseudomodes with $W$ given by~\eqref{eqn:ND2W}.}
    (b) The simplest non-diagonalizable pseudomode configuration\edit{, with $\zeta_2=\eta=0$}. One mode is coupled to the system and the other is coupled to the residual bath. If the energies and couplings are chosen so that $W$ is non-diagonalizable, then the system feels an effective spectral density which is a squared Lorentzian, as in~\eqref{eqn:squared-lorentzian}.}
    \label{fig:2layermeso}
\end{figure}

Consider the case where we map bath $\alpha$, which couples to only one system site (so the spectral density matrix is $1\times1$) to two pseudomodes, so $W_\alpha$ is $2\times2$. 
If we want $W_\alpha$ to be non-diagonalizable, the most general form we can pick is:
\begin{equation}\label{eqn:ND2W}
W_\alpha=\begin{bmatrix}
-i\varepsilon-\eta & -i\delta\\
-i\delta & -i\varepsilon-\eta-2\delta
\end{bmatrix}\,,
\end{equation}
with system couplings ${\zeta=\begin{bmatrix}\zeta_1 & \zeta_2\end{bmatrix}^T}$. 
\edit{This corresponds to the configuration shown in Fig.~\ref{fig:2layermeso}(a).}
The effective memory kernel~\eqref{eqn:zeWtz} then has an exponential component and a component that is $t$ times an exponential:
\begin{equation}
\begin{aligned}
\chi_\alpha^{\rm eff}(t)=e^{-i\varepsilon t}e^{-(\delta+\eta)t}\Big\{\left(|\zeta_1|^2+|\zeta_2|^2\right)+\\\delta t\left[|\zeta_1|^2-|\zeta_2|^2-i\left(\zeta_1\zeta_2^*+\zeta_1^*\zeta_2\right)\right]
\Big\}\,.
\end{aligned}\end{equation}
The resulting effective spectral density~\eqref{eqn:Jeff} then has three terms: a Lorentzian, an Anti-Lorentzian, and a term that looks like a squared Lorentzian plus a squared Anti-Lorentzian:
\begin{align}
    \label{eqn:ND2J}
    J^{\rm eff}_\alpha(\omega)=\left(\left|\zeta_1\right|^2+\left|\zeta_2\right|^2\right)\frac{2(\delta+\eta)}{(\omega-\varepsilon)^2+(\delta+\eta)^2}\nonumber\\[0.2cm]
    +(\zeta_1\zeta_2^*+\zeta_1^*\zeta_2)\frac{4\delta(\delta+\eta)(\omega-\varepsilon)}{\left[(\omega-\varepsilon)^2+(\delta+\eta)^2\right]^2}\\[0.2cm]
    +\left(\left|\zeta_1\right|^2-\left|\zeta_2\right|^2\right)
    \frac{2\delta\left[(\delta+\eta)^2-(\omega-\varepsilon)^2\right]}{\left[(\omega-\varepsilon)^2+(\delta+\eta)^2\right]^2}\nonumber\,.
\end{align}

If we take the case where $\eta=0$ and $\zeta_2=0$, then this $W$ represents two pseudomodes in a chain-like configuration, with the system coupled to mode $1$, which is then coupled to mode $2$, which is coupled to a residual bath, as shown in Fig.~\ref{fig:2layermeso}(b). 
In this case, the spectral density we obtain is a squared Lorentzian:
\begin{equation}
    \label{eqn:squared-lorentzian}
    J^{\rm eff}_\alpha(\omega)=\frac{4\delta^3|\zeta_1^2|}{\left[(\omega-\varepsilon)^2+\delta^2\right]^2}\,.
\end{equation}

We can generalize this to the case where we have several pairs of pseudomodes, where each pair has its own $\varepsilon_k, \delta_k, \zeta_{k}, \eta_k$ and obtain a spectral density that is a sum of terms of the form~\eqref{eqn:ND2J}.

These non-diagonalizable configurations allow for more variety in the types of effective spectral densities that can be obtained from the pseudomode method beyond only Lorentzians and Anti-Lorentzians, though the terms obtained this way can be closely approximated by just Lorentzians and Anti-Lorentzians.
If we treat a non-diagonalizable configuration as the limit of a barely-diagonalizable configuration, then the resulting diagonalizable-type spectral density approaches a non-diagonalizable one, as shown in Appendix~\ref{apx:NDasLimit}.
However, these non-diagonalizable configurations also provide clues into how to fit more general types of spectral densities. 
For example, from Table~\ref{tab:ndterms} we can see that a spectral density whose denominator contains terms of order up to $\omega^6$ might be able to be fit exactly by just 3 modes. \edit{An additional example of a non-diagonalizable configuration is shown in Appendix~\ref{apx:ND3}.}

\section{Few-Mode Fitting\label{sec:fitting}}
While determining the effective spectral density due to a particular pseudomode configuration is fairly straightforward, what we generally want to do is the reverse process: start with a bath with a known spectral density and determine the pseudomode parameters that best model that bath.
Specifically, we wish to choose the pseudomode parameters with the following goals:
\begin{enumerate}[label=(\roman*)]
    \item The effective spectral density matrix $\Jeff(\omega)$ should closely match the true spectral density matrix $\Jaij(\omega)$.
    \item The number of pseudomodes used should be as low as possible to make the problem more amenable to computer simulations.
\end{enumerate}

Depending on how strongly we prioritize (ii), multiple approaches exist.
While most approaches prioritize using a small number of modes, if instead we are willing to use a large number of pseudomodes, one option is to proceed similarly to Refs.~\cite{brenes2020a, lacerda2023b, lacerda2024, bettmann2025}  as described in section~\ref{sec:mesofitting}. 
In the remainder of this section, we focus on few-mode fitting.
We will assume a diagonalizable configuration for this section where not otherwise specified. 
Thus, the effective memory kernel is given by~\eqref{eqn:Xeff_diagonalizable} and the effective spectral density is given by~\eqref{eqn:Jeff_diagonalizable}.

The most common approach to few-mode fitting is to first fit the memory kernel in terms of the parameters $\kappa_{\alpha k,ij},\varepsilon_{\alpha k},\gamma_{\alpha k}$, and then to try to choose pseudomode parameters $\Lambda,\Gamma,\zeta$ that give rise to the same effective memory kernel. 
We focus on this method.
\edit{One might alternately choose to fit the correlation functions~\eqref{eqn:correlations_basic} (or their Fourier transforms), thus encoding the temperatures and chemical potentials of the original baths into the pseudomode parameters, rather than the residual baths~\cite{pleasance2020a,menczel2024a}.}
Few-mode fitting has also been performed by brute-force optimization over the pseudomode parameters without
first fitting the memory kernel~\cite{feist2020,medina2021,sanchez-barquilla2022}; doing so guarantees
a fit with physical pseudomode parameters \editt{(i.e, $\Gamma>0$)}, which we shall see is not always guaranteed otherwise, but is much less
efficient computationally. 
A third method was recently proposed in Ref.~\cite{huang2025}, which instead \edit{maps~\eqref{eqn:Jeff} onto the so-called control-theoretic realization problem~\cite{mayo2007, antoulas1986}}\editt{ and also guarantees a fit with physical pseudomode parameters}.

In the remainder of this section, we present a new method for choosing the pseudomode parameters, for which the only optimization step is the fitting of the memory kernel. Using this method, the pseudomode parameters can be chosen to exactly match the $\kappa_{\alpha k,ij},\varepsilon_{\alpha k},\gamma_{\alpha k}$ obtained by fitting the memory kernel.

\subsection{Fitting the Memory Kernel\label{ssec:Xfit}}
We fit the memory kernel from each bath separately. 
We first fit each element of $\xeff(t)$ as a sum of complex exponentials using techniques from discrete signal processing, such as Prony's method~\cite{deprony1795,wilson2019,almunif2020,chen2022} or the ESPRIT algorithm~\cite{roy1986,paulraj1986,sahnoun2017} (see Appendix~\ref{apx:prony}).  
\edit{These methods work by sampling the true $\chi(t)$ at discrete, evenly spaced times, and then fitting those discrete points with complex exponentials.
We found the best results by using ESPRIT, as it is more stable with respect to the set of points at which $\chi$ is sampled than Prony's method.}
This gives us the set of $\kappa_{\alpha k,ij},\varepsilon_{\alpha k},\gamma_{\alpha k}$ from which to construct the effective memory kernel~\eqref{eqn:Xeff_diagonalizable}.
Since the exponents in $\xeff$ are independent of $i,j$, in the end each $i,j$ element of the memory kernel matrix should be fit by a sum of exponentials with the same exponents but different amplitudes $\kappa_{\alpha k,ij}$.
Caution must be used here, since the fit obtained may produce a spectral density that is not positive everywhere or even has non-decaying exponentials.
Note, also, that some spectral densities will be easier to fit than others. 
In particular, if we have a wideband $J(\omega)=\Gamma_0$ for $\omega_m\leq\omega\leq\omega_M$ and $0$ otherwise, then the corresponding memory kernel is 
\begin{equation}
\chi(t)=\frac{i\Gamma_0}{t}\left(e^{-i\omega_M t}-e^{-i\omega_m t}\right)\,.
\end{equation}

Because this memory kernel decays as $1/t$, it cannot be well-fit by any finite set of decaying exponentials.
Thus, somewhat ironically, a large but flat spectral density is a very difficult problem to describe with the pseudomode approach. 

\subsection{Obtaining Pseudomode Parameters\label{ssec:inversion}}
Choosing the parameters $\Lambda,\Gamma,\zeta$ to obtain the desired amplitudes $\kappa$ is an Inversion Problem, and it remains the greatest challenge in implementing the pseudomode technique. 
The relation~\eqref{eqn:kappa} between the constants $\kappa_{\alpha k,ij}$ and the actual pseudomode parameters $\Lambda_{\alpha},\Gamma_{\alpha},\zeta_{\alpha}$ is highly nonlinear, and there are many more degrees of freedom in these parameters than are required. 
The fit with Prony's method (or similar algorithms) for $n$ pseudomodes gives us $4n$ real parameters: $n$ complex amplitudes and $n$ complex exponents. 
For the actual pseudomode parameters, we have $n^2+3n$ real free parameters: $n^2$ from $\Lambda$ (Hermitian), $n$ from $\Gamma$ (real and diagonal), and $2n$ from $\zeta$ (complex).
However, not all of these free parameters actually contribute due to the structure of the $\kappa$'s~\eqref{eqn:kappa}, and there are a number of constraints that are not readily apparent. 
For example, a simple one is that, because we always see pairs $\zeta_{\alpha i}^*\zeta_{\alpha j}^{}$, the vector $\zeta_{\alpha i}$ is invariant under a global phase $\zeta\to e^{i\phi}\zeta$.
\edit{Similarly, we can express the pseudomode operators $a_{\alpha i}$ in any basis without changing the dynamics. We choose the basis in which $\Gamma$ is diagonal so that the resulting master equation will have local dissipators. Physically, this is equivalent to the specification in~\eqref{eqn:H_Meso} and~\eqref{eqn:flat-residuals} that each pseudomode couples to its own residual bath and that each of those residual baths has a constant spectral density $\Gamma_{\alpha,kk}$.}

We can formally express this Inversion Problem as follows. 
For this section we will suppress the indices $i,j$ and consider the case where each bath only couples to one system site; we will also suppress the bath index $\alpha$ since each bath is fit independently.
Given a fit consisting of a set of $\kappa_k,\varepsilon_k,\gamma_k$, our goal is to construct $W$ and $\zeta$. Now,
\begin{equation}\label{eqn:WandM}
    W=SMS^{-1}\,,\quad M_{ij}=\delta_{ij}(-i\varepsilon_i-\gamma_i/2)\,,
\end{equation}
where $M$ is a diagonal matrix whose entries are the eigenvalues of $W$ obtained from fitting the memory kernel. 
While the eigenvalues of $W$ are specified, $S$ is not.
Note that if we transform $W$ by some unitary transformation, then as long as we transform $\zeta$ by the same unitary we will obtain the same effective memory kernel~\eqref{eqn:Xeff_diagonalizable}:
\begin{equation}\begin{aligned}
    \chi^{\rm eff}_{ij}(t)&=(U\zeta_i)^\dag US e^{Mt} S^{-1}U^\dag(U\zeta_j)\\
    &=\zeta_i^\dag Se^{Mt}S^{-1}\zeta_j\,.
\end{aligned}\end{equation}
We must then choose the elements of the matrix $S$ and vector $\zeta$ to get the right amplitudes $\kappa$ according to~\eqref{eqn:kappa}. 

We can rewrite the equation for the $\kappa_k$ in a more useful form as:
\begin{equation}
\begin{aligned}
    \kappa_k&=(\zeta^\dag S)_k(S^{-1}\zeta)_k\\
    &=(v^\dag)_ku_k\,,
\end{aligned}
\end{equation}
where we define the vectors
\begin{align}\label{eqn:simple_uv}
    u=S^{-1}\zeta\,,\qquad v=S^\dag \zeta.
\end{align}
Note that the relation between these two vectors is
\begin{equation}\label{eqn:u=STSv}
    S^\dag Su=v\,.
\end{equation}
Thus, we can express the Inversion Problem as the problem of, given a vector $\kappa_k$, choosing two vectors $u,v$ such that
\begin{equation}\label{eqn:TheBIP}
    v_k^* u_k=\kappa_k
\end{equation}
which are related by a positive semi-definite matrix $S^\dag S$. Notice that $u^\dag v=\sum_k\kappa_k>0$ is guaranteed from~\eqref{eqn:kappa_sum}.
Equation~\eqref{eqn:TheBIP} is now bi-linear in $u,v$. 
We have thus expressed this as a Bi-linear Inversion Problem (BIP), a class of problems that tend to be highly underdetermined~\cite{li2015d,aghasi2018}. 

We can write any solution to~\eqref{eqn:TheBIP} by first fixing the values of $u_j$\edit{, which must be nonzero,} and then taking
\begin{equation}\label{eqn:vfromu}
    v_j=(\kappa_j/u_j)^*\,.
\end{equation}

Our task is now to construct a matrix $S^\dag S$ which is positive definite and satisfies~\eqref{eqn:u=STSv}. We can see such a matrix for $n$ pseudomodes will have the form
\begin{equation}\label{eqn:vuPlusBperp}
    S^\dag S=\frac{vu^\dag}{||u||^2}+B_\perp\,,
\end{equation}
where $B_\perp$ can be any positive matrix that satisfies $B_\perp u=0$ and makes~\eqref{eqn:vuPlusBperp} Hermitian. 
\edit{The choice of $B_\perp$ matrix will not affect the effective spectral density, but will affect the pseudomode parameters $\Lambda,\Gamma$. In particular, this will affect the rates in the master equation, which are given by the diagonal entries of $\Gamma$.}

We will use a convenient change of basis to construct a matrix of this form. 
For $n$ pseudomodes, we choose any vectors $b_1,...,b_{n-2}$ such that the set $\{v,u,b_1,...,b_{n-2}\}$ is linearly independent. 
We then perform Graham-Schmidt orthonormalization on this set to obtain a new orthonormal basis $\{e_1,e_2,...,e_n\}$. 
In this basis, note that by construction we have
\begin{equation}
    v=||v||e_1\,,\quad {u=u_1 e_1+u_2e_2}\,,
\end{equation}
where $u$ and $v$ only have one or two nonzero elements. The unitary matrix that implements this change of basis is the matrix whose columns are the $e_j$:
\begin{equation}
    U_e=\begin{bmatrix}
        \vert&\vert&&\vert\\
        e_1&e_2&\dots&e_n\\
        \vert&\vert&&\vert
    \end{bmatrix}\,.
\end{equation}
Thus, we can construct the matrix $S^\dag S$ as block-diagonal in the $e_j$ basis with two blocks
\begin{equation}\label{eqn:STS_e}
    S^\dag S=U_e\begin{bmatrix}
        A&\mathbf{0}\\
        \mathbf{0}&B
    \end{bmatrix}U_e^\dag\,,
\end{equation}
where the block $A$ is $2\times 2$ and satisfies
\begin{equation}\begin{aligned}\label{eqn:A_block}
    &A_{11}u_1+A_{12}u_2=||v||\,,\\
    &A_{12}^*u_1+A_{22}u_2=0\,,
\end{aligned}\end{equation}
and the block $B$ is $(n-2)\times(n-2)$ and positive definite. 
This is analogous to the decomposition in~\eqref{eqn:vuPlusBperp}.
A simple choice is to take $B=\mathds{1}_{n-2}$ as the identity matrix. 
The matrix $A$ is then constructed by solving~\eqref{eqn:A_block} for $A_{11},A_{22}$; we then choose $A_{12}$ so that $S^\dag S$ is positive semi-definite, which requires
\begin{equation}\label{eqn:A12}
    |A_{12}|^2<A_{11}A_{22}\,.
\end{equation}

Once we have constructed the blocks $A,B$, we can then obtain $S^\dag S$ from~\eqref{eqn:STS_e}.
Since $S^\dag S>0$, its square root $S=\sqrt{S^\dag S}$ exists, is well-defined, and is Hermitian. 
From there, we use~\eqref{eqn:simple_uv} and~\eqref{eqn:WandM} to construct
\begin{equation}\label{eqn:W_tilde}
    \tilde{\zeta}=Su\,,\quad\tilde{W}=SMS^{-1}\,,\quad\tilde\Gamma=-\left(\tilde W+\tilde W^\dag\right)\,.
\end{equation}

In general, the matrix $\tilde\Gamma$ will not be diagonal at this stage. Thus, we define $U_\Gamma$ as the unitary matrix that diagonalizes $\tilde\Gamma$
\begin{equation}
    \Gamma=U_\Gamma^{}\tilde\Gamma U_\Gamma^\dag\,,
\end{equation}
and then rotate both $\zeta$ and $W$ by this same unitary:
\begin{gather}
    W =U_\Gamma^{}\tilde{W}U_\Gamma^\dag=U_\Gamma^{} SMS^{-1}U_\Gamma^\dag\,,\\
    \zeta=U_\Gamma\tilde\zeta=U_\Gamma Su\,.
\end{gather}

We can now finally construct the actual pseudomode parameters $\Lambda,\Gamma,\zeta$ corresponding to a set of $\kappa_k,\varepsilon_k,\gamma_k$ obtained from Prony's method:
\begin{gather}\label{eqn:inversion_solution}
    \Lambda=\frac{1}{2i}(W^\dag-W)\,,\quad\Gamma=-(W+W^\dag)\,.
\end{gather}

This approach can always be used to obtain a set of pseudomode parameters whose effective spectral density will exactly match the effective spectral density obtained from fitting the memory kernel, provided that the fitting procedure used outputs a set of $\kappa_k$ that satisfy~\eqref{eqn:kappa_sum}. However, this approach is \textit{not} guaranteed to give a matrix $\Gamma>0$.
\begin{figure*}[!t] 
    \centering
    \includegraphics[width=0.49\linewidth]{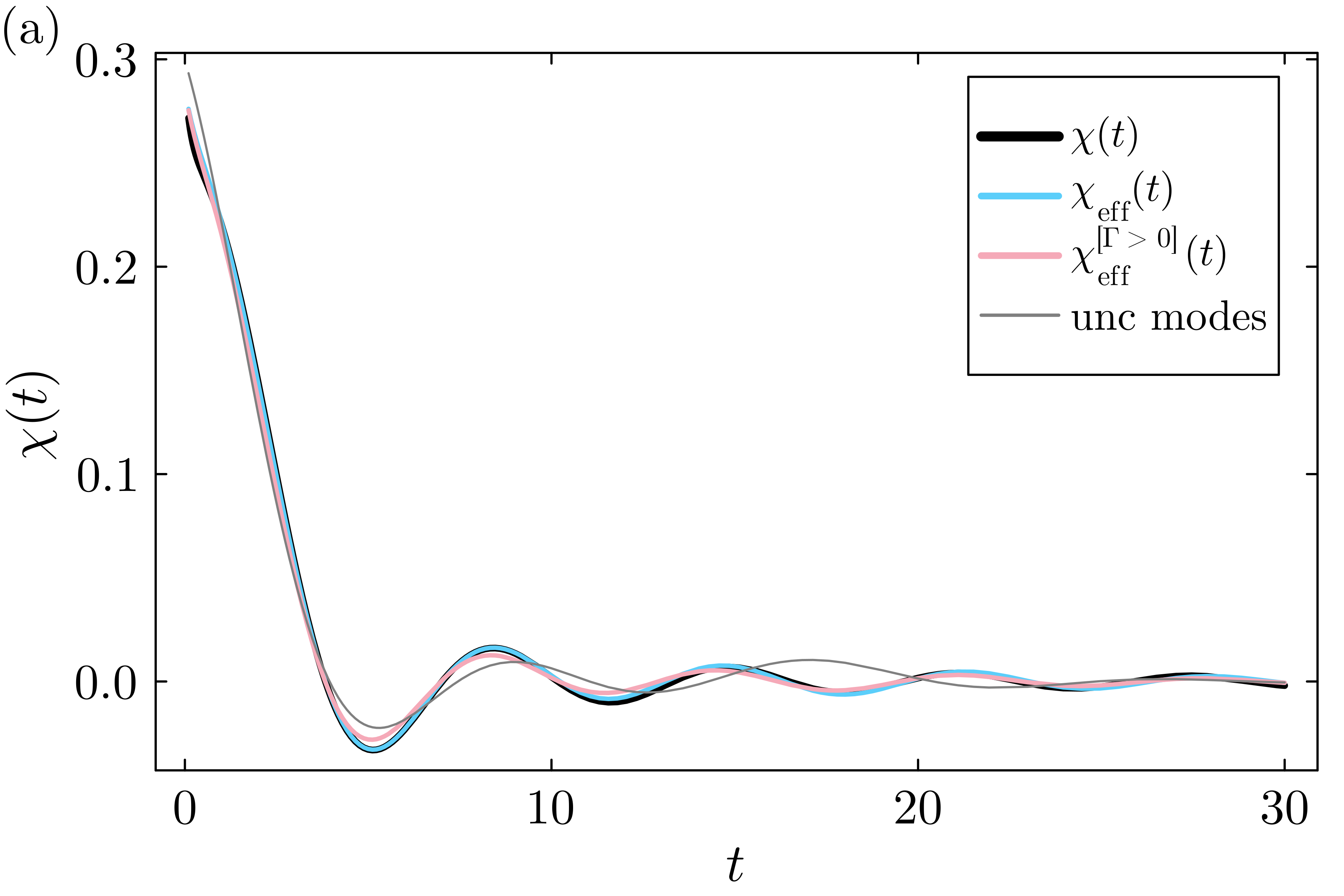}
    \includegraphics[width=0.49\linewidth]{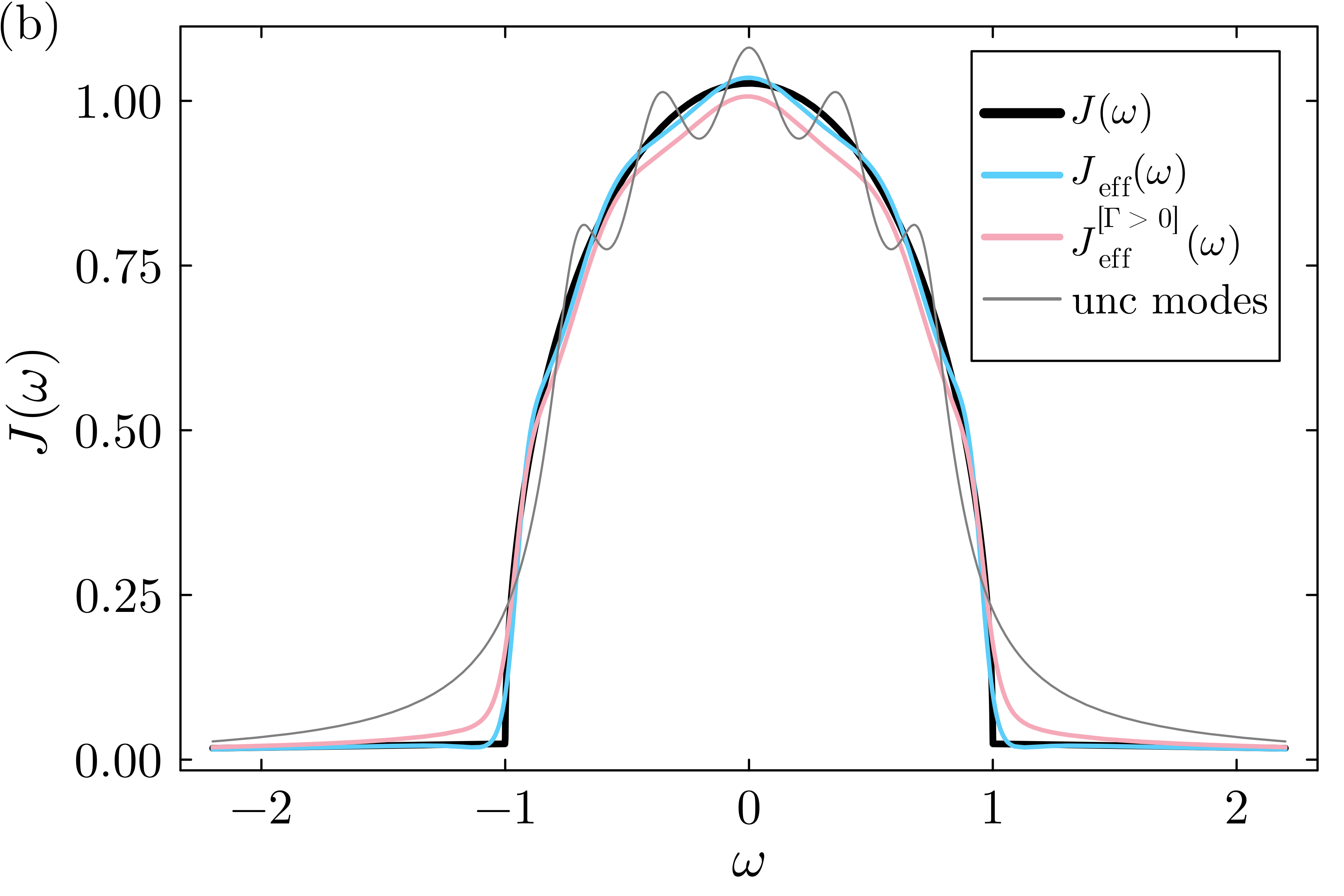}
    \caption{\edit{Fits to (a) a memory kernel of the form~\eqref{eqn:numex-X} and (b) the corresponding spectral density~\eqref{eqn:numex-j}, with $\Delta=0.04$ and $\delta=6.0$. The blue curves are obtained by fitting~\eqref{eqn:numex-X} with the ESPRIT algorithm (see Appendix~\ref{apx:prony}). After following the inversion procedure of section~\ref{ssec:inversion}, optimizing over the vector $u$~\eqref{eqn:vfromu} and matrix $B$~\eqref{eqn:STS_e} to minimize any negative elements of $\Gamma$, one obtains the pseudomode parameters $\Lambda,\zeta,\Gamma$ corresponding to that fit. Finally, setting any negative elements of the matrix $\Gamma$ (which would cause negative rates in the corresponding master equation) gives rise to the modified fit shown in red. These fits are compared with the best fit using uncoupled modes, which was obtained using brute-force optimization. The pseudomode parameters appearing in these fits are given in Appendix~\ref{apx:data}.
    }}
    \label{fig:numex}
\end{figure*}

It is worth emphasizing again here the sheer amount of freedom involved in this process. In~\eqref{eqn:vfromu} we can choose anything for the $u_j$. 
Then in~\eqref{eqn:vuPlusBperp}, $B_\perp$ can by \textit{any} positive-definite $n\times n$ matrix satisfying $B_\perp u=0$; alternately, we can choose any positive block $B$ in~\eqref{eqn:STS_e} and any value of $A_{12}$ satisfying~\eqref{eqn:A12}. 
A naturally interesting question is then as follows: For what fit parameters $\kappa_k,\varepsilon_k,\gamma_k$ will there be a choice of $u_j,B_\perp$ such that we obtain $\Gamma>0$? 
We were unable to find a general answer to this question, but we explore it further for the simplest case of two pseudomodes in Appendix~\ref{apx:2x2inversion}. For that case, we show that it is possible to have $\Gamma>0$ if and only if the effective spectral density from the fit parameters is positive everywhere.

\subsection{\edit{Numerical Implementation}}
\edit{
As an example, we now apply this inversion procedure to a spectral density which is the sum of a semicircular term with a wide Lorentzian:
\begin{equation}\label{eqn:numex-j}
    J(\omega)=\sqrt{(1-\omega)^2}\,+\frac{\Delta\delta}{\omega^2+(\delta/2)^2}\,.
\end{equation}
We choose this spectral density because fermionic system coupled to an infinite, homogeneous wire will have a semi-elliptical spectral density~\cite{ferracin2024b}. 
We add the wide Lorentzian to represent how no physical system has true hard cutoff frequencies in its spectral density, and because the methods of section~\ref{ssec:Xfit} struggle to fit the sharp corners of a semicircular spectral density without producing an effective spectral density that is negative somewhere.}

\edit{
The corresponding memory kernel is
\begin{equation}\label{eqn:numex-X}
    \chi(t)=\frac{1}{2t}\mathcal{J}_1(t)+\Delta e^{-\delta t}
\end{equation}
where $\mathcal{J}_1(t)$ is a Bessel function. We first fit $\chi(t)$ with pseudomodes using the ESPRIT algorithm~\cite{paulraj1986,roy1986,sahnoun2017} (see Appendix~\ref{apx:prony}). Then we optimize over the values of the vector $u$ from~\eqref{eqn:vfromu} and the matrix $B$ from~\eqref{eqn:STS_e}, trying first to make any negative rates as small as possible, and then minimizing the largest rate if $\Gamma>0$.
}

\edit{As an example, we construct a six-mode fit to the spectral density~\eqref{eqn:numex-j}, using $\Delta=0.04$ and $\delta=6.0$. When optimizing over $u,B$, we obtain five positive rates and one negative rate. 
The spectral density and memory kernel and their corresponding fits are shown in Figs.~\ref{fig:numex}(a) and~\ref{fig:numex}(b) respectively, along with the effective fits that we obtain if we set the one negative rate to zero. 
We compare our results with an optimized fit involving only uncoupled modes. While in this case we are unable to obtain a fit with only nonnegative rates, our modified fit where the negative rates are discarded still fits the true memory kernel quite well at both short and long times with only minor deviations at the peaks and troughs of $\chi(t)$ at intermediate times.
}

\section{Many-Mode Fitting \label{sec:mesofitting}}
If we are willing to use a large number of pseudomodes, we can ensure that couplings to the residual baths are arbitrarily small. 
However, it stops being practical to run optimization algorithms for fitting. 
In this case, we might proceed as in Refs.~\cite{brenes2020a, lacerda2023b, lacerda2024, bettmann2025}. 
The procedure uses a diagonal configuration, so each pseudomode couples only to the system and to its own residual bath and the effective spectral density is given by~\eqref{eqn:Jeff_diagonal}. 
There are a few benefits to this many-mode procedure. 
It is more efficient to set up since no optimization routine is required, though computation with a large number of modes is expensive. 
Additionally, by using a large number of uncoupled modes at different energies, analyzing the currents between individual pseudomodes and the system can provide information about energy currents. 
For each bath, the procedure is as follows, where the index $\alpha$ is suppressed:
\begin{enumerate}
    \item Define a relevant frequency range $[\omega_{m},\omega_{M}]$ over which to approximate the spectral density.
    \item Choose a number of pseudomodes $n$ such that \begin{equation}\gamma=\frac{\omega_{M}-\omega_{m}}{d-1}\end{equation} is an acceptably small coupling to the residual baths.
    \item Let $\varepsilon_j=\omega_m+(j-1)\gamma$, such that the $\varepsilon_j$ form an evenly spaced array with $\varepsilon_1=\omega_m$ and $\varepsilon_d=\omega_M$.
    \item Construct $\Lambda,\Gamma$ by $\Lambda_{ij}=\delta_{ij}\varepsilon_j$ and $\Gamma_{ij}=\delta_{ij}\gamma$.
    \item Choose the elements of $\zeta$ so that\\ $\zeta_{ i,k}\zeta_{j,k}=J_{ij}(\varepsilon_k)\gamma/2\pi$.
\end{enumerate}

Since the matrices $\Lambda,\Gamma$ are diagonal, the effective spectral density matrix is given by~\eqref{eqn:Jeff_diagonal}. Furthermore, since all the $\zeta_{ik}$ are real, the resulting elements of the spectral density matrix $J_{ij}$ will be sums of Lorentzians centered at the elements $\varepsilon_k$ with width $\gamma/2$:
\begin{equation}\label{eqn:MesoJeff}
    \Jxff(\omega)=\sum_{k=1}^{n}\frac{J_{ij}(\varepsilon_k)}{2\pi}\frac{\gamma^2}{\left(\omega-\varepsilon_k\right)^2+(\gamma/2)^2}\,.
\end{equation}

\begin{figure}[!bt] 
    \centering
    \includegraphics[width=\linewidth]{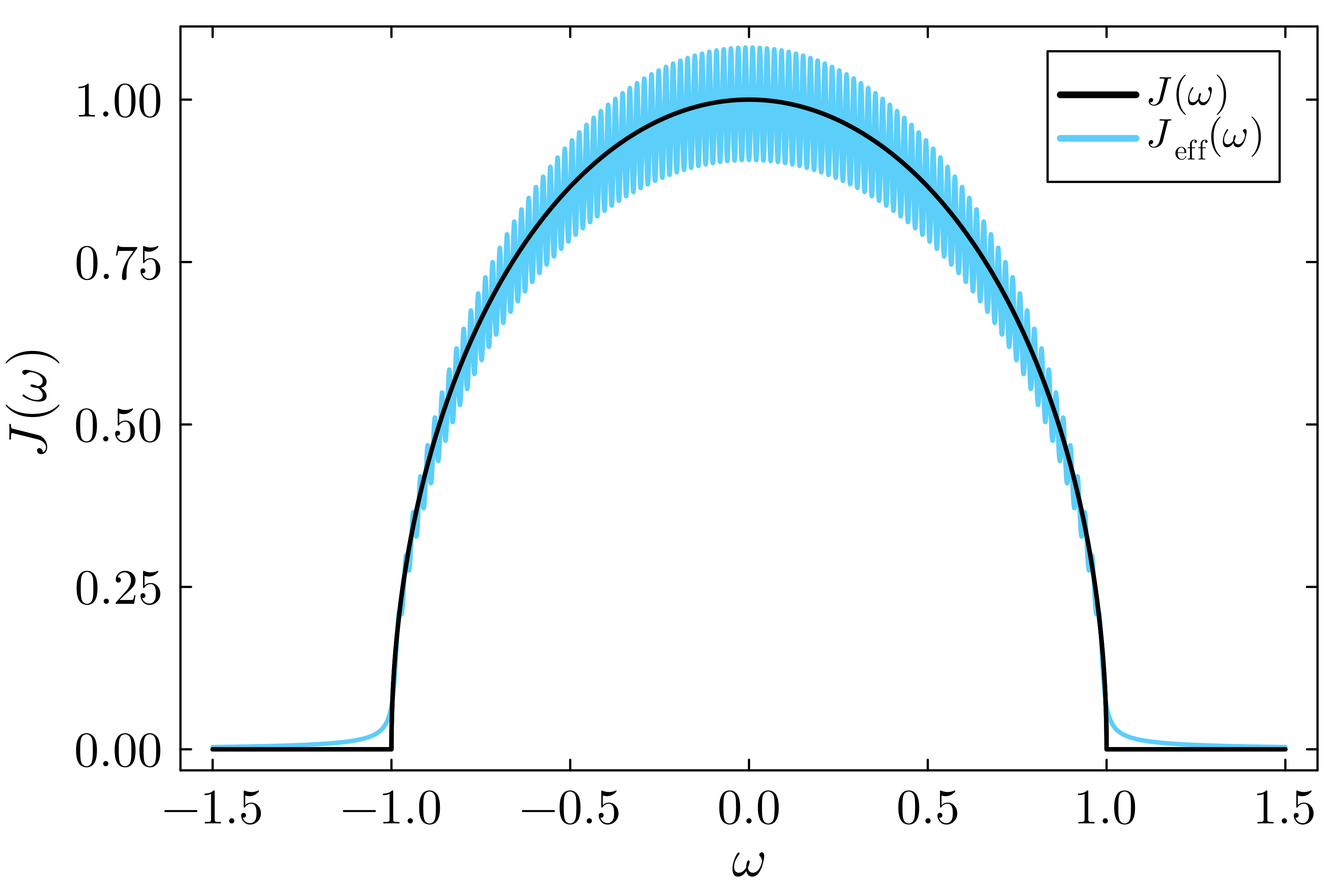}
    \caption{Effective spectral density obtained by applying the many-mode approximation procedure described above to a semi-elliptical spectral density, for 100 pseudomodes with $\omega_{m/M}=\pm 1.4$. The effective spectral density oscillates rapidly about the true spectral density.}
    \label{fig:nonconv}
\end{figure}

It has been claimed that this sum will converge to $\Jaij(\omega)$ as $n\to\infty$ (assuming we adjust $\gamma$ accordingly); however, this is not the case.
Directly applying the procedure described above for a large number of modes to a semi-elliptical spectral density yields the effective spectral density depicted in Fig.~\ref{fig:nonconv}, which oscillates rapidly about the true spectral density. 

Note that the problem of non-convergence presented here arises from this particular method of choosing the pseudomode parameters. In general, one can always choose parameters to approximate a spectral density arbitrarily well, even in the diagonal (uncoupled) case~\cite{trivedi2021a}.
 
\subsection{Proof of Nonconvergence}
The problem of nonconvergence arises because the spacing between Lorentzians is also used as the width of the Lorentzians, and so the ratio between their widths and separations remains constant rather than tending to zero and we cannot reduce the Lorentzians to delta functions.

To show where the issue emerges, consider $\Jxff(\varepsilon_k)$, the effective spectral density at one of the pseudomode energies. 
\edit{Since $\gamma$ is defined as the spacing between pseudomode energies $(\varepsilon_k-\varepsilon_{k-1})$,} we can write\edit{~\eqref{eqn:MesoJeff} evaluated at $\varepsilon_k$ as}:
\begin{equation}\label{eqn:nonconv-ekei}
\Jxff(\varepsilon_k)=\sum_{q=1}^n\frac{J_{ij}(\varepsilon_q)}{2\pi\left[(q-k)^2+1/4\right]}\,,
\end{equation}
where the $\gamma$'s conveniently cancel since $\varepsilon_q - \varepsilon_k = (q-k)\gamma$. 
Now, assuming a very large number of pseudomodes, we can approximate $\Jxff(\varepsilon_q)$ in~\eqref{eqn:nonconv-ekei} by 
\begin{equation}
    \label{eqn:nonconv-taylor}
    \Jxff(\varepsilon_q)\approx\Jxff(\varepsilon_k)+J'(\varepsilon_k)(q-k)\gamma\,,
\end{equation} 
where the shorthand $J'=dJ/d\omega$. 
\edit{This approximation is reasonable in the case that we have a large number of pseudomodes, since terms in the sum where $|q-k|$ is large are suppressed quadratically.}
We then can see that the contributions from the linear terms at $\varepsilon_{k+q}$ and $\varepsilon_{k-q}$ will cancel each other, so we can rewrite~\eqref{eqn:nonconv-ekei} as:
\begin{equation}
\label{eqn:nonconv-ekek}
\Jxff(\varepsilon_k)=\sum_{q\edit{=1}}^{\edit{n}}\frac{J_{ij}(\varepsilon_k)}{2\pi\left[(q-k)^2+1/4\right]}\,.
\end{equation}

We can arrive at~\eqref{eqn:nonconv-ekek} more rigorously if we restrict the sum to a set number of terms ($q$ such that $|k-q|\leq 8$ is sufficient to show nonconvergence) and use the fact that $J_{ij}$ is a continuous function to show that we can always choose a number of modes so that $J_{ij}(\varepsilon_k)-J_{ij}(\varepsilon_q)<\epsilon$ for any $\epsilon$. 
In the limit of infinitely many pseudomodes we can take $\epsilon\to0$ and the limits on the summation to $\pm\infty$ and likewise arrive at~\eqref{eqn:nonconv-ekek}.
\edit{This scheme works even at points where the linear approximation~\eqref{eqn:nonconv-taylor} is poor, since it only relies on the spectral density being continuous.}

In the limit where the number of pseudomodes goes to infinity, we can then reindex the summation in~\eqref{eqn:nonconv-ekek} as:
\begin{equation}\label{eqn:nonconv-l}
    \Jxff(\varepsilon_k)=\frac{J_{ij}(\varepsilon_k)}{2\pi}\sum_{\ell=-\infty}^\infty\frac{1}{\ell^2+1/4}\,,
\end{equation}
and finally evaluating the summation:
\begin{equation}
\Jxff(\varepsilon_k)=J_{ij}(\varepsilon_k)\coth(\pi/2)\approx1.09J_{ij}(\varepsilon_k)\,.
\end{equation} 
\begin{figure*}[!tb]
    \centering
    \includegraphics[width=0.49\linewidth]{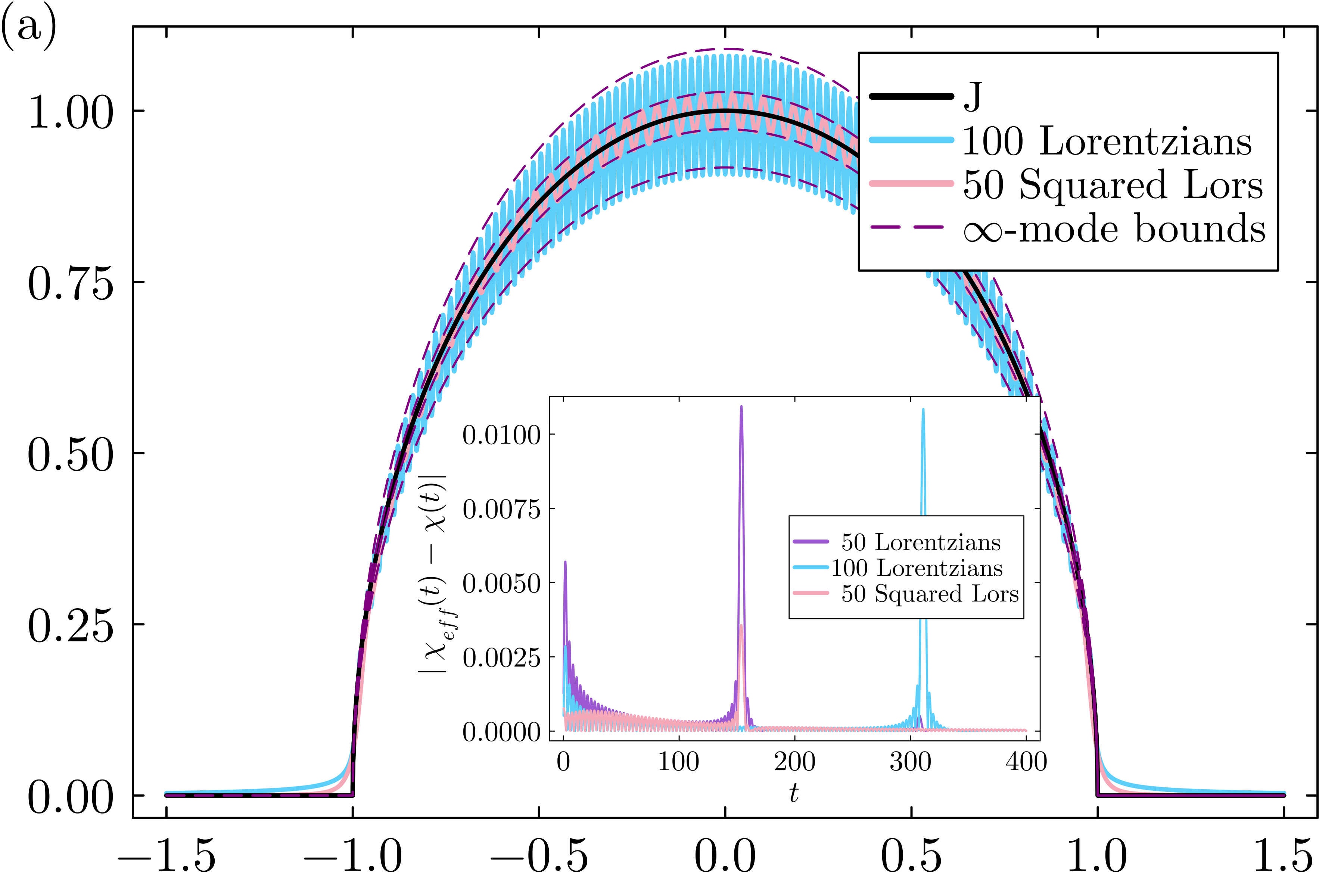}
    \includegraphics[width=0.49\linewidth]{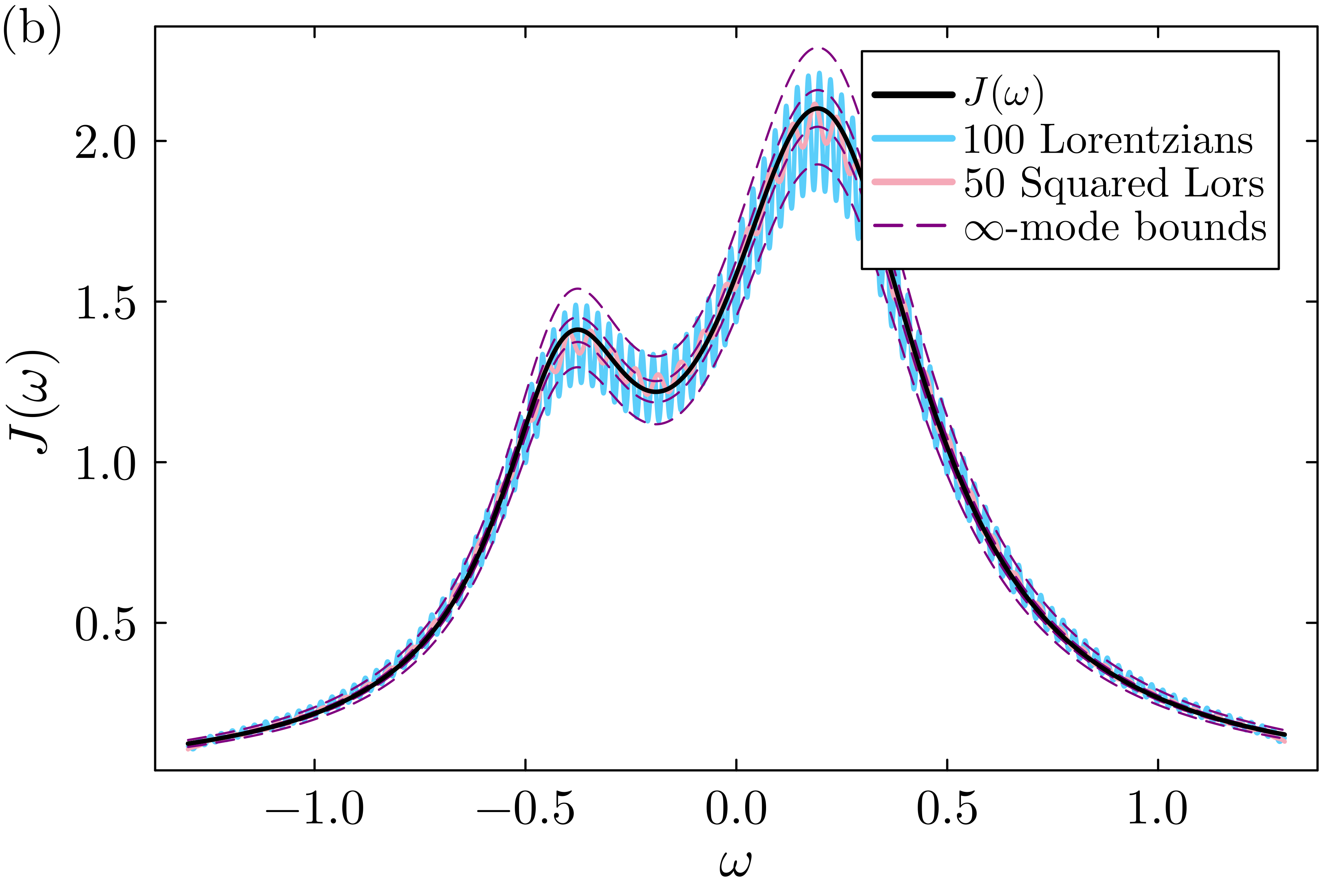}
    \caption{Effective spectral density for pseudomodes in a diagonal configuration (blue) and for modes in $2\times2$ non-diagonalizable blocks (pink), compared to the error expected in the infinite-mode limit at the pseudomode energies (above $J_{\rm eff}$) and halfway between them (below $J_{\rm eff}$), for two example spectral densities. 
    100 pseudomodes are used in each \edit{fit in each} figure, creating an effective spectral density that is either the sum of 100 Lorentzians \edit{(100 uncoupled modes)} or 50 squared Lorentzians \edit{(100 modes, coupled as 50 pairs)}, with $\omega_{m,M}=\pm 1.0$ for both figures. 
    Note that where the effective spectral density differs from the infinite-mode prediction, it is where the second derivative of the true spectral density is large, so the first-order Taylor approximation~\eqref{eqn:nonconv-taylor} is locally a poor approximation. The left spectral density is semi-elliptical, and the right is a sum of two Lorentzians.
    \editt{The inset to (a) depicts the corresponding error in the time domain for the fits depicted in the main figure, as well as for a fit involving 50 uncoupled modes. The rapidly oscillating error in frequency space corresponds to a sharp peak at a large time in time space.}}
    \label{fig:convergence-bounds}
\end{figure*}

We can use this same process to find the complete effective spectral density in the infinite-mode limit with this same process. 
If we let $r$ be between $0$ and $1$ then we can write $\Jxff(\varepsilon_j+r\gamma)=\eta_1(r)J_{ij}(\varepsilon_j+r\gamma)$ where using the same technique to reach an infinite summation we now have:
\begin{align}
\label{eqn:eta1}
    \eta_1(r)&=\frac{1}{2 \pi }\left({\sum _{\ell=0}^{\infty } \frac{1}{(\ell+r)^2+\frac{1}{4}}}+\sum _{\ell=1}^{\infty } \frac{1}{ (\ell-r)^2+\frac{1}{4}}\right)
    \\&=\frac{-\sinh (\pi )}{\cos (2 \pi  r)-\cosh (\pi )}\,,
\end{align}
which we can see is periodic in $r$ with period 1. 
We can see these oscillations in Fig.~\ref{fig:convergence-bounds}.

\subsection{Many-Mode Approach with Non-diagonalizability}
We can do slightly better in the large-$n$ limit at the cost of doubling the number of modes by replacing the pseudomodes above with $2\times2$ non-diagonalizable blocks, with one of the two pseudomodes in each block coupled to the system and the other coupled to the residual bath as shown in Fig.~\ref{fig:2layermeso}, such that each block contributes a squared Lorentzian to the effective spectral density as in~\eqref{eqn:squared-lorentzian}. 
The steps above are modified as:
\begin{enumerate}
    \item Define a relevant frequency range $[\omega_{m},\omega_{M}]$ over which to approximate the spectral density
    \item Choose an even number of pseudomodes $n$ such that $\gamma=4\left(\frac{\omega_{M}-\omega_{m}}{d/2-1}\right)$ is an acceptably small coupling to the residual baths \edit{(with the factor of $4$ included because the squared Lorentzians have width $\delta=\gamma/4$).}
    \item Let $\delta=\gamma/4$ and $\varepsilon_{q}=\left[\omega_{m}+\delta(q-1)\right]$ for ${q=1,...,n/2}$, such that the $\varepsilon_q$ form an evenly spaced array with ${\varepsilon_1=\omega_m}$ and ${\varepsilon_{d/2}=\omega_M}$.
    \item Construct $\Lambda$ and $\Gamma$ as block-diagonal, with $n/2$ $2\times2$ blocks $\Lambda^q$ and $\Gamma^q$ of the form:
    \begin{equation}
        \Lambda^q=\begin{bmatrix}
            \varepsilon_q & \delta \\ \delta & \varepsilon_q
        \end{bmatrix}\,,\qquad\quad
        \Gamma^q=\begin{bmatrix}
            0&0 \\ 0&4\delta
        \end{bmatrix}\,.
    \end{equation}
    \item Choose the elements of $\zeta$ so that\\ $\zeta_{ i,k}\zeta_{j,k}=J_{ij}(\varepsilon_{(k+1)/2})\gamma/2\pi$ if $k$ is odd and 0 if $k$ is even.
\end{enumerate}

Under this method, we obtain an effective spectral density that is a sum of $n/2$ squared Lorentzians:
\begin{equation}
    J^{\rm eff}_{ij}(\omega)=\sum_{k=1}^{n/2} \frac{4 J_{ij}(\varepsilon_{k})}{2\pi} \frac{\delta^4}{\left[(\omega-\varepsilon_{k})^2+\delta^2\right]^2}\,.
\end{equation}
There is still error in the large-$n$ limit for this method, but it is smaller than that for the one-layer version. 
If we proceed analogously to before, we obtain at the pseudomode energies
\begin{align}
    J^{\rm eff}_{ij}(\varepsilon_k)&= \frac{2J_{ij}(\varepsilon_k)}{\pi}\sum_{\ell=-\infty}^\infty \frac{1}{(\ell^2+1)^2}\\
    &=\frac{2}{\pi}\big[\coth(\pi)+\pi\mathrm{csch}(\pi)\big]\Jaij(\varepsilon_k)\\
    &\approx 1.027J_{ij}(\varepsilon_k)\,,
\end{align}
and in general we obtain $J^{\rm eff}_{ij}(\varepsilon_k+r\gamma)=\eta_2(r)J_{ij}(\varepsilon_k+r\gamma)$ with
\begin{align}
\label{eqn:eta2}
    \eta_2(r)&=\frac{2}{\pi}{\left(\sum _{\ell=0}^{\infty } \frac{1}{\left[ (\ell+r)^2+1\right]^2}+\sum _{\ell=1}^{\infty } \frac{1}{\left[ (\ell-r)^2+1\right]^2}\right)}\\[0.2cm]
    &=\frac{\cos (2 \pi  r) (4 \pi  \cosh (2 \pi )-2 \sinh (2 \pi ))-4 \pi +\sinh (4 \pi )}{2 (\cos (2 \pi  r)-\cosh (2 \pi ))^2}\,.
\end{align}
We can numerically compare the two methods to see these differences in error, as shown in Fig.~\ref{fig:convergence-bounds}. \editt{This rapidly oscillating error in the frequency domain becomes, in the time domain, a large sharply-peaked error around some time $T$ which increases with the number of pseudomodes used, as shown in the inset to Fig.~\ref{fig:convergence-bounds}(a).}
\edit{This oscillating error can be further mitigated by increasing $\gamma$ and adjusting the elements of $\zeta$ accordingly. However, while this reduces the magnitude of the oscillations, it makes the approximation~\eqref{eqn:nonconv-taylor} (and the corresponding approximation in the two-layer case) worse, and fails to fit the spectral density anywhere its second derivative is too large. It is likely that this issue could be mitigated by further refining this many-mode fitting procedure.}

\section{Connection to Scattering Theory\label{sec:scattering}}
\edit{The method of pseudomodes is typically used to treat interacting systems. 
However, in this section we turn our attention to the methods of non-equillibrium Green's functions (NEGF) and scattering theory, a common method for non-interacting systems~\cite{blanter2000a, buttiker1992a, levitov1996a, esposito2009b}, to show how the same pseudomode structure and effective spectral density that we derived using quantum master equations emerges in the context of scattering theory as well.
The results of this section are largely independent of the previous section. They have been included here for two reasons. 
First, they illustrate how the same pseudomode structure can emerge from an entirely different approach. 
Second, they allow one to compute transmission functions resolved in the different lead modes. This makes it possible to study the tunneling amplitudes between the two baths, through the system, but resolved in the different lead modes, possibly providing new insights into the problem. 
} 

If the system Hamiltonian in~\eqref{eqn:H_basic} is non-interacting, we can fully describe the steady-state transport properties by specifying the transmission probability for an electron with energy $\omega$ to tunnel from bath $\beta$ to bath $\alpha$, which is given by~\cite{blanter2000a}:
\begin{equation}\label{T_basic}
    T_{\alpha\beta} (\omega) = \Tr\big\{ J_\alpha(\omega) G(\omega) J_\beta(\omega) G^\dagger(\omega)\big\}\,.
\end{equation}
Here $\alpha,\beta \in \{L,R\}$, so this includes both transmission and reflection. 
The quantity $G$ is the single-particle delayed non-equilibrium Green's function (NEGF), which is an $n_S\times n_S$ matrix given by 
\begin{equation}\label{G_basic}
    G(\omega) = \frac{1}{\omega - H_S - \Sigma(\omega)}\,,
\end{equation}
where $\Sigma(\omega)= \sum_\alpha \Sigma_\alpha(\omega)$ is the self-energy
\begin{equation}\label{Sigma_basic}
    \Sigma_\alpha(\omega) = -\frac{i J_\alpha(\omega)}{2} + 
    \Upsilon_\alpha(\omega)\,,
    \quad 
    \Upsilon_\alpha(\omega):=
    \mathcal{P} \int\frac{d\Omega}{2\pi} \frac{J_\alpha(\Omega)}{\omega-\Omega}\,.
\end{equation}
The imaginary part is the spectral density and the real part $\Upsilon_\alpha$ is the level shift.

Under the pseudomode mapping with Hamiltonian~\eqref{eqn:H_Meso}, we treat the system and the two set of pseudomodes as an extended system. 
We can then study the transmission probability for an electron with energy $\omega$ to tunnel between  residual bath $(\alpha, k)$ and residual bath $(\beta, q)$. 
This therefore adds an additional layer of resolution to the tunneling probabilities. 
To do that, we first recall that the effects of the residual baths in each pseudomode are captured by the spectral densities:
\begin{equation}\label{ML_J}
    J_{\alpha k}^{\rm res}(\omega) = 2\pi \sum_{p\edit{\in(\alpha,k) }} \left|\tau_{\alpha k, p}\right|^2 \delta(\omega- \omega_p)\,,
\end{equation}
as well as the corresponding level shifts $\Upsilon_{\alpha k}^{\rm res}$, defined as in~\eqref{Sigma_basic}. 
Here, because each pseudomode is coupled to its own reservoir,~\eqref{ML_J} is a scalar, rather than a matrix like~\eqref{eqn:J_basic}.

We can now use a formula akin to~\eqref{T_basic} to write the transmission probability $T_{\alpha k, \beta q}^{\rm res}$ between residual baths. 
To do that,  introduce the NEGF of the extended system:
\begin{equation}\label{ML_G}
    \mathcal{G}(\omega) = \frac{1}{\omega- Q(\omega)}\,,
\end{equation}
where 
\begin{equation}\label{ML_Q}
    Q(\omega) = \begin{bmatrix}
        \Lambda_L + \Sigma_L(\omega) & 0 & \zeta_L \\
        0 & \Lambda_R + \Sigma_R(\omega) & \zeta_R \\
        \zeta_L^\dagger & \zeta_R^\dagger & H_S
    \end{bmatrix}\,,
\end{equation}
which is of size $n_L+n_R+n_S$. 
The matrices $\zeta_\alpha$ are of size $n_\alpha \times n_S$. 
The matrices $\Lambda_\alpha$ are the same ones that appear in~\eqref{eqn:H_Meso}. 
And the matrices $\Sigma_\alpha$ are diagonal, with entries $-iJ_{\alpha k}^{\rm res}/2 + \Upsilon_{\alpha k}^{\rm res}$ in each diagonal. 

The transmission probabilities will then have the form 
\begin{equation}\label{ML_T}
    T_{\alpha k, \beta q}^{\rm res} (\omega) = 
    \Tr\big\{ \mathcal{X}_{\alpha k}(\omega) \mathcal{G}(\omega) \mathcal{X}_{\beta q}(\omega) \mathcal{G}^\dagger(\omega)\big\}\,.
\end{equation}
Once again, this describes transmissions from left to right, as well as reflections. 
Moreover, it also describes reflections to the same pseudomode, as well as to the different pseudomodes belonging to the same bath. 
The matrices $\mathcal{X}_{\alpha k}$ in~\eqref{ML_T} are diagonal, of size $d_L+d_R+n_S$,  with a single non-zero entry $J_{\alpha k}^{\rm res}$ at position $(\alpha,k)$ (in the same block structure as~\eqref{ML_Q}).

Our next goal is to express~\eqref{ML_T} in terms of matrices with dimension $n_S$ of the system alone. This will allow us to compare the results with those of~\eqref{T_basic}. 
Consider  the Green's function~\eqref{ML_G} and  parametrize it in blocks as 
\begin{equation}\label{ML_G_block}
    \mathcal{G} = \begin{bmatrix}
        \mathcal{G}_{LL} & \mathcal{G}_{LR} & \mathcal{G}_{LS} \\
        \mathcal{G}_{RL} & \mathcal{G}_{RR} & \mathcal{G}_{RS} \\ 
        \mathcal{G}_{SL} & \mathcal{G}_{SR} & \mathcal{G}_S
    \end{bmatrix}\,.
\end{equation}
Using the fact that the matrices $\mathcal{X}_{\alpha k}$ in~\eqref{ML_T} are all sparse, we can write four cases of interest: 
\begin{align}
\label{TLL}
    T_{L k, L q}^{\rm res}  &=   ~\Tr \big\{ 
    \hat{J}_{L k}^{\rm res}\mathcal{G}_{LL}\hat{J}_{L q}^{\rm res}\mathcal{G}_{LL}^\dagger\big\}\,,
    \\[0.2cm]
\label{TRR}    
    T_{Rk, Rq}^{\rm res} &=  ~\Tr \big\{ \hat{J}_{R k}^{\rm res}\mathcal{G}_{RR}\hat{J}_{R q}^{\rm res} \mathcal{G}_{RR}^\dagger\big\}\,,
    \\[0.2cm]
\label{TLR}    
    T_{Lk, Rq}^{\rm res} &=   ~\Tr \big\{ \hat{J}_{L k}^{\rm res}\mathcal{G}_{LR}\hat{J}_{R q}^{\rm res}\mathcal{G}_{LR}^\dagger\big\}\,,
    \\[0.2cm]
\label{TRL}
    T_{Rk, Lq}^{\rm res} &=   ~\Tr \big\{ \hat{J}_{R k}^{\rm res}\mathcal{G}_{RL}\hat{J}_{L q}^{\rm res}\mathcal{G}_{RL}^\dagger\big\}\,,
\end{align}
where the dependence on $\omega$ has been omitted for clarity. 
Here $\hat{J}_{\alpha k}^{\rm res}$ is a matrix with dimension $n_\alpha \times n_\alpha$ having only one entry $J_{\alpha k}^{\rm res}$ in position $k$ of the diagonal. 
These results can also be written more compactly as 
\begin{equation}\label{T_alpha_beta}
    T_{\alpha k, \beta q}^{\rm res} = \Tr\big\{ \hat{J}_{\alpha k}^{\rm res} \mathcal{G}_{\alpha \beta} \hat{J}_{\beta q}^{\rm res} \mathcal{G}_{\alpha \beta}^\dagger\big\}\,.
\end{equation}
To proceed, we therefore need the blocks $\mathcal{G}_{LL}, \mathcal{G}_{RR}, \mathcal{G}_{LR}, \mathcal{G}_{RL}$, which we will express in terms of $\mathcal{G}_S$, the effective system NEGF (the (3,3) entry in~\eqref{ML_G_block}).

Equation~\eqref{ML_G} means that $\mathcal{G}$ is the solution of the equations 
\begin{align}
    (\omega-Q)\mathcal{G} &= 1,
    \\
    \mathcal{G}(\omega-Q) &= 1. 
\end{align}
Using the block forms in~\eqref{ML_G_block} and~\eqref{ML_Q} this yields the following system of coupled equations. 
First, for the (1,1) and (2,2) diagonal entries we get  
\begin{align}\label{diag_12_12}
    (\omega-\Lambda_\alpha-\Sigma_\alpha) \mathcal{G}_{\alpha\alpha} &= 1+ \zeta_\alpha \mathcal{G}_{S\alpha}\,, 
    \\
    \mathcal{G}_{\alpha\alpha} (\omega-\Lambda_\alpha-\Sigma_\alpha)  &= 1+ \mathcal{G}_{\alpha S} \zeta_\alpha^\dagger\,.
\end{align}
Next, for the (3,3) entries: 
\begin{align}
    \label{ne1h18h19811}
    (\omega- H_S)\mathcal{G}_S &= 1 +  \big\{ \zeta_L^\dagger \mathcal{G}_{LS} + \zeta_R^\dagger \mathcal{G}_{RS}\big\}\,, 
    \\
    \mathcal{G}_S(\omega- H_S) &= 1 +  \big\{ \mathcal{G}_{SL}\zeta_L + \mathcal{G}_{SR} \zeta_R\big\}\,.
\end{align}
Finally, for the off-diagonal entries we get: 
\begin{align}
    (1,2): \quad \label{GLR}
    \mathcal{G}_{LR} &= \mathcal{F}_L \zeta_L \mathcal{G}_{SR}= \mathcal{G}_{LS} \zeta_R^\dagger \mathcal{F}_L
    \\[0.2cm]
    (2,1): \quad \label{GRL}
    \mathcal{G}_{RL} &= \mathcal{F}_L \zeta_R \mathcal{G}_{SL}= \mathcal{G}_{RS} \zeta_L^\dagger \mathcal{F}_L
    \\[0.2cm]
    (1,3): \quad \label{j91081n91}
    \mathcal{G}_{LS} &= \mathcal{F}_L \zeta_L \mathcal{G}_S= \big\{ \mathcal{G}_{LL} \zeta_L + \mathcal{G}_{LR} \zeta_R \big\} \frac{1}{\omega-H_S}
    \\[0.2cm]
    (3,1): \quad 
    \mathcal{G}_{SL} &= \frac{1}{\omega-H_S} \big\{ \zeta_L^\dagger \mathcal{G}_{LL} + \zeta_R^\dagger \mathcal{G}_{RL}\big\}= \mathcal{G}_S \zeta_L^\dagger \mathcal{F}_L
    \label{GSL2}
    \\[0.2cm]
    (2,3): \quad \label{h910u1j98181}
    \mathcal{G}_{RS} &= \mathcal{F}_L \zeta_R \mathcal{G}_S= \big\{ \mathcal{G}_{RL} \zeta_L + \mathcal{G}_{RR} \zeta_R \big\} \frac{1}{\omega-H_S}
    \\[0.2cm]
    (3,2): \quad 
    \mathcal{G}_{SR} &= \frac{1}{\omega-H_S} \big\{ \zeta_L^\dagger \mathcal{G}_{LR} + \zeta_R^\dagger \mathcal{G}_{RR} \big\}= \mathcal{G}_S \zeta_R^\dagger \mathcal{F}_L
    \label{GSR_2}
\end{align}
where we defined 
\begin{equation}
    \mathcal{F}_\alpha = \frac{1}{\omega- \Lambda_\alpha - \Sigma_\alpha}\,.
\end{equation}
Plug~\eqref{j91081n91} and~\eqref{h910u1j98181} 
into~\eqref{ne1h18h19811} to find 
\begin{equation}
    \left(\omega - H_S - \zeta_L^\dagger \mathcal{F}_L \zeta_L  - \zeta_R^\dagger \mathcal{F}_L \zeta_R\right) \mathcal{G}_S = 1\,.
\end{equation}
Hence, the effective system NEGF is given by 
\begin{equation}\label{GS_eff}
    \mathcal{G}_S = \frac{1}{\omega - H_S - \sum_\alpha \zeta_\alpha^\dagger \mathcal{F}_\alpha(\omega) \zeta_\alpha}\,.
\end{equation}

Next we express the four quantities $\mathcal{G}_{LL}, \mathcal{G}_{RR}, \mathcal{G}_{LR}, \mathcal{G}_{RL}$ [which appear in~\eqref{TLL}–\eqref{TRL}] in terms of $\mathcal{G}_S$.
Combining~\eqref{GLR} with~\eqref{GSR_2} and~\eqref{GRL} with~\eqref{GSL2} we get 
\begin{equation}
    \mathcal{G}_{LR} = \mathcal{F}_L \zeta_L \mathcal{G}_S \zeta_R^\dagger \mathcal{F}_L\,,\quad\mathcal{G}_{RL} = \mathcal{F}_L \zeta_R \mathcal{G}_S \zeta_L^\dagger \mathcal{F}_L\,.
\end{equation}
Finally, combining~\eqref{diag_12_12} with~\eqref{GSL2} or~\eqref{GSR_2} we get 
\begin{equation}
    \mathcal{G}_{\alpha\alpha} = \mathcal{F}_\alpha + \mathcal{F}_\alpha \zeta_\alpha \mathcal{G}_S \zeta_\alpha^\dagger \mathcal{F}_\alpha\,.
\end{equation}
In short, we conclude that 
\begin{equation}\label{ML_NEGF_connections}
    \mathcal{G}_{\alpha \beta}  = \delta_{\alpha \beta} \mathcal{F}_\alpha + \mathcal{F}_\alpha \zeta_\alpha \mathcal{G}_S \zeta_\beta^\dagger \mathcal{F}_\beta\,,
\end{equation}
which is the general formula connecting all relevant NEGFs to the system NEGF.

As the last step, we plug~\eqref{ML_NEGF_connections} into~\eqref{T_alpha_beta}. 
First, for $\beta\neq \alpha$ we get 
\begin{equation}
\begin{aligned}
    T_{\alpha k, \beta q}^{\rm res} &=
    \Tr\big\{ \hat{J}_{\alpha k}^{\rm res} (\mathcal{F}_\alpha \zeta_\alpha \mathcal{G}_S \zeta_\beta^\dagger \mathcal{F}_\beta) 
    \hat{J}_{\beta q}^{\rm res} (\mathcal{F}_\beta^\dagger \zeta_\beta \mathcal{G}_S^\dagger \zeta_\alpha^\dagger \mathcal{F}_\alpha^\dagger)\big\}
    \\[0.2cm]
    &=\Tr\big\{ \big( \zeta_\alpha^\dagger \mathcal{F}_\alpha^\dagger \hat{J}_{\alpha k}^{\rm res} \mathcal{F}_\alpha \zeta_\alpha) \mathcal{G}_S (\zeta_\beta^\dagger \mathcal{F}_\beta \hat{J}_{\beta q}^{\rm res} \mathcal{F}_\beta^\dagger \zeta_\beta)\mathcal{G}_S^\dagger\big\}\,.
\end{aligned}    
\end{equation}
We therefore see that we can write this in a form similar to~\eqref{T_basic}, provided we define $n_S\times n_S$ matrices
\begin{equation}\label{eqn:curlyJ}
    \mathcal{J}_{\alpha k}^+ = -\zeta_\alpha^\dagger \mathcal{F}_\alpha^\dagger \hat{J}_{\alpha k}^{\rm res} \mathcal{F}_\alpha \zeta_\alpha\,, 
    \qquad 
    \mathcal{J}_{\alpha k}^- = -\zeta_\alpha^\dagger \mathcal{F}_\alpha \hat{J}_{\alpha k}^{\rm res} \mathcal{F}_\alpha^\dagger \zeta_\alpha\,.
\end{equation}
Then, finally: 
\begin{equation}
    T_{\alpha k, \beta q}^{\rm res} = \Tr\big\{ \mathcal{J}_{\alpha k}^+ \mathcal{G}_S \mathcal{J}_{\beta q}^- \mathcal{G}_S^\dagger\big\}\,.
\end{equation}

\edit{As in~\eqref{eqn:flat-residuals}}, we now take the pseudomodes' residual spectral densities to be constant: 
\begin{equation}\label{eqn:flatresidual}
    J^{\rm res}_{\alpha k}(\omega)=\Gamma_{\alpha\edit{,kk}}\,.
\end{equation}
Then, if we let $\mathcal{J}_\alpha^+(\omega)=\sum_k\mathcal{J}_{\alpha k}(\omega)$, we can see that, starting from~\eqref{eqn:curlyJ} with $W_\alpha$ as in~\eqref{eqn:W_matrix}:
\begin{align}
    \mathcal{J}^\pm_{\alpha}(\omega)&=-\zeta_\alpha^\dag\frac{1}{\omega+iW^\dag_\alpha}\Gamma_\alpha\frac{1}{\omega-iW_\alpha}\zeta_\alpha\\[0.2cm]
    &=\zeta_\alpha^\dag\frac{1}{\omega+iW_\alpha^\dag}(W_\alpha+W_\alpha^\dag)\frac{1}{\omega-iW_\alpha}\zeta_\alpha\\[0.2cm]
    &=i\zeta^\dag\left(\frac{1}{iW_\alpha-\omega}+\frac{1}{iW_\alpha^\dag+\omega}\right)\zeta_\alpha\\[0.2cm]
    &=2\Im\left(\zeta_\alpha^\dag\frac{1}{iW_\alpha-\omega}\zeta_\alpha\right)\,.
\end{align}
So as per~\eqref{eqn:Jeff} we have that
\begin{equation}
    \sum_k\left[\mathcal{J}_{\alpha k}^\pm(\omega)\right]_{ij}=\Jeff(\omega)\,.
\end{equation}

Likewise, if we have constant residual spectral densities~\eqref{eqn:flatresidual}, then ${\Sigma_\alpha=i\Gamma_\alpha/2}$, and so the effective system Green's function is:
\begin{equation}
    \mathcal{G}_S(\omega)=\frac{1}{\omega-H_S-\sum_\alpha\zeta_\alpha^\dag\frac{1}{\omega+iW_\alpha}\zeta}\,.
\end{equation}

Recall that the true Green's function is:
\begin{equation}\label{eqn:Geff-almost}
    G(\omega)=\frac{1}{\omega-H_S+\sum_\alpha\frac{iJ_\alpha(\omega)}{2}-\mathcal{P}\int\frac{d\Omega}{2\pi}\frac{J_\alpha(\Omega)}{\omega-\Omega}}\,.
\end{equation}

The principal value $\Upsilon_\alpha$ in the NEGF arises from forcing the memory kernel to be causal:
\begin{equation}\label{eqn:Greal-almost}
    \int dt\,\chi_\alpha(t)e^{i\omega t}\theta(t)=\frac{J_\alpha(\omega)}{2}-\mathcal{P}\int\frac{d\Omega}{2\pi}\frac{J_\alpha(\Omega)}{\omega-\Omega}\,.
\end{equation}
where $\theta(t)$ is the Heaviside step function.
We can see the equivalence between~\eqref{eqn:Geff-almost} and~\eqref{eqn:Greal-almost} by noticing that
\begin{equation}
    \int dt\,\xeff(t)e^{i\omega t}\theta(t)=\int dt\,\zeta_\alpha^\dag e^{(W_\alpha+i\omega)t}\zeta_\alpha=\zeta_\alpha\frac{1}{W_\alpha+i\Omega}\zeta_\alpha\,.
\end{equation}

We can then write in the pseudomode picture:
\begin{gather}\label{eqn:T_eff}
    T^{\rm eff}_{\alpha\beta}(\omega)=\Tr\left\{J^{\rm eff}_\alpha(\omega)
    \mathcal{G}_S(\omega)J^{\rm eff}_\beta(\omega)\mathcal{G}_S^\dag(\omega)\right\}\,,\\
    \mathcal{G}_S(\omega)=\frac{1}{\omega-H_S+\sum_\alpha \frac{iJ_\alpha^{\rm eff}(\omega)}{2}-\mathcal{P}\int\frac{d\Omega}{2\pi}\frac{J^{\rm eff}_\alpha(\Omega)}{\omega-\Omega}}\,,
\end{gather}
where $\mathcal{G}_S$ is just the NEGF that we would calculate from the system interacting with a set of baths with spectral densities $J^{\rm eff}_\alpha$. 
Thus, we can see that so long as the true and effective spectral densities (or correlation functions) match exactly, the transmission functions~\eqref{T_basic} and~\eqref{eqn:T_eff} will also match exactly.

\section{Conclusion\label{sec:conclusion}}
The method of pseudomodes is a useful approach to studying the dynamics of open quantum systems with strong coupling or structured environments.
We have explored a number of subtleties that arise in the implementation of this technique, including the differences in effective spectral densities for coupled and uncoupled pseudomodes and the appearance of new terms in the effective spectral density when the pseudomodes' non-Hermitian Hamiltonian is non-diagonalizable.
We have explored the pseudomode inversion problem and presented a method for finding pseudomode parameters that exactly match an effective fit to a spectral density.
Our inversion method also sheds further light on the vast number of free parameters involved in the pseudomode method.
We have also shown a connection between pseudomodes and scattering theory for non-interacting systems and shown that they produce the same effective spectral densities.

While the method we present for inversion is exact, it is not guaranteed to provide a pseudomode configuration with positive damping rates. 
We have explored the constraints on obtaining positive rates in the limited case of 2 pseudomodes, and shown that it is possible in that case to obtain positive rates if and only if the effective spectral density is positive everywhere. However, it is currently unclear whether this result will always hold in general.

\section*{Acknowledgments}
This research is primarily supported by the U.S. Department of Energy (DOE), Office of Science, Basic Energy Sciences (BES) under Award No. DE-SC0025516. L.P.B. acknowledges Research Ireland for support through the Frontiers for the Future project. Optimization was performed with the Optim.jl Julia package \cite{mogensen2018}. Numerical data used in this article are provided in Appendix~\ref{apx:data}. No generative AI tools were used in the writing of this article.

\appendix

\section{Derivation of the Effective Spectral Density \label{apx:Jeffderivation}}

In this appendix we derive the forms of the effective spectral density and effective memory kernel given in~\eqref{eqn:Jeff} and~\eqref{eqn:zeWtz}, respectively, following Ref.~\cite{ferracin2024b}. Consider the system described in section~\ref{ssec:mesomodel} with Hamiltonian given by~\eqref{eqn:H_Meso} in the special case where the residual baths have constant spectral densities $J^{\rm res}_{\alpha k}(\omega)=\Gamma_{\alpha k}$ and the Fermi functions of the original (physical) baths are $f_\alpha$. We then model the residual baths using a local master equation of the GKSL form. The Hamiltonian is now:
\begin{equation}
\begin{aligned}
    H = H_S(t) + 
    \sum_{\alpha \in \{L,R\}} \sum_{k,q}  \Lambda_{\alpha, kq} a_{\alpha k}^\dagger a_{\alpha q}^{}
     +\\ \sum_{\alpha \in \{L,R\}} \sum_{i,k} \big(\zeta_{\alpha i,k} a_{\alpha k}^\dagger c_i^{} + \zeta_{\alpha i,k}^* c_i^\dagger a_{\alpha k}^{}\big)\,,
\end{aligned}\end{equation}
such that the total time evolution for an operator $\mathcal{O}$ in the evolving-operators picture is:
\begin{equation}\label{eqn:GKSL}\begin{aligned}
    \frac{d\mathcal{O}}{dt}&=\mathcal{L}'\mathcal{O}=i[H,\mathcal{O}]+\\&\sum_{\alpha\in\{L,R\}}\sum_k \Gamma_{\alpha k}\left(f_{\alpha k}D'[a_{\alpha k}^\dag]\mathcal{O}+(1-f_{\alpha k})D'[a_{\alpha k}]\mathcal{O}
    \right)\,,
\end{aligned}
\end{equation}
where the adjoint dissipator $D'[L]\mathcal{O}=L^\dag\mathcal{O}L-\frac{1}{2}\{L^\dag L,\mathcal{O}\}$.

Since the spectral density and memory kernel are independent of reservoir temperature or chemical potential, we perform the derivation for the special case of a competely empty reservoir: one whose Fermi function $f_\alpha(\omega)=0$. In this special case, the bath correlation function $C_{\alpha ij}^+(t)$ is zero at all times, and $C_{\alpha i j}^-(t)$ is identical to the memory kernel.
\begin{equation}
    \xaij(t)\big|_{f(\omega)=0}=C_{\alpha,ij}^-(t)=\expval{B_{\alpha i}(t)B_{\alpha j}^\dag(0)}\,.
\end{equation}

Additionally, for simplicity, since each bath will have its own independent correlation functions, we consider the case of only one reservoir and so can suppress the index $\alpha$. We can then evaluate the memory kernel matrix $\chi(t)$ by way of the expectation value in the evolving-operators picture~\eqref{eqn:C-expectation}, where $\ketemp$ is the full vacuum state of the bath:
\begin{align}
 \chi_{ij}(t)&=\expval{B_{i}(t)B_{j}^\dagger(0)}\\
&=\sum_k\sum_l\zeta_{ i,k}\zeta_{ j,l}^*\braemp a_{ k}(t)a_{ l}^\dagger(0)\ketemp\\
&=\sum_k\sum_l\zeta_{ i,k}\zeta_{ i,l}^*\braemp a_{ k}(t)\ketone{l}\,,
\end{align}
where $\ketone{l}=a_{ l}^\dag\ketemp$. Now define a vector $v_{ i}$ such that its components are $v_{ i,j}(t)=\braemp a_{ i}(t) \ketone{j}$ and consider:

\begin{align}\frac{dv_{i,j}}{dt}&=\bra{0}\dot{a}_i(t)\ket{1_j}=\bra{0}\mathcal{L}'{a}_i(t)\ket{1_j}\,,
\end{align}
where $\mathcal{L}'$ is the adjoint Liouvillian that arises from evolution under~\eqref{eqn:GKSL}. The two terms are then, where $\mathcal{D}'$ represents only the dissipator part of~\eqref{eqn:GKSL}:

\begin{align}
\braemp[H,a_i(t)]\ketone{j}&=0+\sum_{m=1}^{n_\alpha}\Lambda_{mj}\braemp a_i(t)\ketone{m}\,,\\
\braemp\mathcal{D}'(a_i(t))\ket{1_j}&=-\frac{1}{2}\sum_{m=1}^{n_\alpha}\Gamma_{mj}\braemp a_i(t)\ketone{m}\,.
\end{align}

So combining these and putting them into vector form, we have:
\begin{equation}
    \frac{dv_i}{dt}=\left(-i\Lambda-\frac{1}{2}\Gamma\right)^Tv^i(t)\,.
\end{equation}

If we let $W_{\alpha}=\left(-i\Lambda_\alpha-\frac{1}{2}\Gamma_\alpha\right)$, noting that $iW$ is the non-Hermitian Hamiltonian for this model, then this equation has the solution:
\begin{equation}
    v_{i,j}(t)=\left(e^{tW}\right)_{ij}\,.
\end{equation}

Finally, the memory kernel is then given by the matrix equation:
\begin{equation}
    \chi_{ij}(t)=\zeta_{i}^\dagger e^{Wt}\zeta_{j}\,,
\end{equation}
as given by~\eqref{eqn:zeWtz}. Taking the Fourier transform yields the result for the spectral density given in~\eqref{eqn:Jeff}.
\begin{equation}
    J_{\alpha ij}(\omega)=2\Im\left(\zeta_{\alpha i}^\dag\frac{1}{iW-\omega}\zeta_{\alpha j}\right)\,.
\end{equation}

\section{Memory Kernel Fitting Methods\label{apx:prony}}
Suppose we have a memory kernel $\chi(t)$ that we want to approximate as:
\begin{equation}\label{eqn:apx-tofit}
    \chi(t)=\sum_{k=1}^n \kappa_k e^{-\lambda_k t}\,,
\end{equation}
for complex parameters $\kappa_k,\lambda_k$, where we want $n$ as small as possible to get a good fit. In this appendix we describe two algorithms from discrete signal processing, Prony's method and ESPRIT, that have been used for this fitting problem.

\subsection{Prony's Method}
Prony's method~\cite{deprony1795,wilson2019,almunif2020,chen2022} consists of the following steps:

\begin{enumerate}
    \item Choose a time interval $\Delta t$ and cutoff time $t_c=2M\Delta t$ to sample the correlation function, where $M\geq 2n+1$.
    \item Let $\chi_j=\chi(j \Delta t)$ for $j=0,1,\dots,2M$ and construct the $(M+1)\times(M+1)$ Hankel matrix $\mathbf{H}_{ij}=\chi_{i+j+1}$.
    \item Compute the Takagi decomposition $H=U\sigma U^T$, with $\sigma$ diagonal and $U$ unitary~\cite{ikramov2012}, noting that we take the transpose $U^T$ rather than the adjoint $U^\dag$. This yields a set of normalized vectors $\mathbf{u}_m$ such that $\mathbf{H}\mathbf{u}_m=\sigma_m\mathbf{u}_m^*$. 
    This is analogous to finding the eigen-decomposition of a Hermitian matrix as $\mathbf{H}\ket{m}=\sigma_m\ket{m}$, but for a complex-symmetric matrix. Order the $\sigma_m$ by descending values of $|\sigma_m|$ and the $\mathbf{u}_m$ accordingly.
    \item Define the polynomial $F(z)=\sum_{j=0}^M \left(\mathbf{u}_n\right)_j z^j$ and obtain its roots $f_0,f_1,\dots f_N$ ordered by increasing modulus.
    \item Of these roots, the first $L$ will have $|w_k|<1$. Write these roots as $r_k e^{i\theta_k}$ with $-\pi<\theta\leq\pi$ and choose the exponents in~\eqref{eqn:apx-tofit} to be 
    \begin{equation}
    \lambda_k=-\frac{2M}{t_c}\left[\log(r_k)+i\theta_k\right]\,.
    \end{equation}
    \item Obtain the amplitudes $\kappa_k$ by least-squares fitting the equations 
    \begin{equation}
    {\chi_j=\sum_{k=1}^n \kappa_kw_k^j}\,,\quad j=0,1,\dots n
    \end{equation}
\end{enumerate}

\subsection{ESPRIT Algorithm}
The ESPRIT (Estimation of Signal Parameters by Rotational Invariance Techniques) algorithm~\cite{paulraj1986,roy1986,sahnoun2017} is designed to be less numerically unstable and more robust to noise than Prony's method. 
It consists of the following steps:

\begin{enumerate}
    \item Choose a time interval $\Delta t$ and cutoff time $t_c=M\Delta t$ to sample the correlation function, where $M\geq n+1$.
    \item Let $\chi_j=\chi\big((j-1) \Delta t\big)$ for $j=1,\dots,2M+1$.
    \item Choose the parameter $n+1\leq L< M$, let $K=M-L+1$, and construct the $L\times K$ Hankel matrix $\mathbf{H}_{ij}=\chi_{i+j-1}$.
    \item Compute the singular value decomposition $H=U\Sigma V$,
    where $U$ is the $L\times L$ matrix of left singular vectors, $V$ is $K\times K$, and $\Sigma$ is the $L\times K$ matrix of singular values.
    \item Define $U_s$ to be the $L\times n$ matrix obtained by taking the $n$ columns of $U$ corresponding to the $n$ largest singular values, \edit{and construct the matrix $F$ by solving
    \begin{equation}
        \underline{U_s}F=\overline{U_s}\,,
    \end{equation}}
    where $\underline{U_s}$ ($\overline{U_s}$) is the $(L-1)\times n$ matrix obtained by taking the matrix $U_s$ and removing the bottom (top) row. \edit{(Note that while $U$ is unitary, the matrices $U_s,\overline{U_s},\underline{U_s}$ are \textit{not} unitary).}
    \item Compute the eigenvalues of $F$, denoted $f_k$. These relate to the exponents in~\eqref{eqn:apx-tofit} by:
    \begin{equation}
        f_k=e^{-\lambda_k\Delta t}\,.
    \end{equation}
    \item Obtain the amplitudes $\kappa_k$ by least-squares fitting the equations 
    \begin{equation}
    \chi_j=\sum_{k=1}^n \kappa_kf_k^j\,,\quad j=0,1,\dots n\,.
    \end{equation}
\end{enumerate}

Fitting with either of these methods can be sensitive to the choice of discretization made in step 1\edit{, though ESPRIT is far more stable than Prony's method and can handle a larger number of samples of $\chi(t)$.}
Thus, it is often practical to sample a large number of interval choices and then select the choice that gives the effective spectral density that best fits the true spectral density, or to perform optimization on the endpoints of the time interval used.

\section{Non-Diagonalizable Configuration as the Limit of a Diagonalizable One \label{apx:NDasLimit}}
Consider the $2\times2$ non-diagonalizable block from section~\ref{sssec:ND2} where we set $\varepsilon=\eta=0$, and apply a small perturbation $\epsilon$ to the coupling between the two modes:

\begin{equation}
    W=\begin{bmatrix}
0&-i\delta(1+\epsilon)\\
-i\delta(1+\epsilon)&-2\delta
\end{bmatrix}\,.
\end{equation}
$W$ is diagonalizable so long as $\epsilon\neq 0$, and its eigenvalues are 
\begin{equation}
    -\delta\left(1\pm i\sqrt{2\epsilon(1+\epsilon)}\right)\,.
\end{equation}
We then compute the memory kernel using~\eqref{eqn:zeWtz} and obtain:
\begin{equation}
\begin{aligned}\chi(t)=e&^{-t\delta} \Bigg[\zeta _1^2 \cos \left(\delta  t \sqrt{\epsilon  (\epsilon +2)}\right)+\zeta _2^2 \cos \left(\delta  t \sqrt{\epsilon  (\epsilon +2)}\right)+\\[0.2cm] 
&\frac{\left(\zeta _1^2-\zeta _2^2-2 i \zeta _2 \zeta _1 (\epsilon +1)\right) \sin \left(\delta  t \sqrt{\epsilon  (\epsilon +2)}\right)}{\sqrt{\epsilon  (\epsilon +2)}}\Bigg]\,.\end{aligned}
\end{equation}

In the limit of $\epsilon\to 0$, the first two terms become ${(\zeta_1^2+\zeta_2^2)e^{-t\delta}}$.
The third term is proportional to $\frac{\sin(x)}{x}$ and so in the limit we get $t\delta e^{-t\delta}(\zeta_1-i\zeta_2)^2$, the signature we would expect from a non-diagonalizable $W$.
When we compute the resulting spectral density, we obtain:

\begin{equation}
    J(\omega)=\frac{4 \delta  \left(\delta  \zeta _1+\zeta _2 \omega \right){}^2}{\left(\delta ^2+\omega ^2\right)^2}\,.
\end{equation}

Note that in the case we take $\zeta_2=0$ we recover exactly the squared Lorentzian that we would expect:
\begin{equation}
    J(\omega)=\frac{4 \delta ^3 \zeta _1^2}{\left(\delta ^2+\omega ^2\right)^2}\,.
\end{equation}

If instead we construct the spectral density \textit{before} taking the limit, then we can define $\kappa_k=\sum _{i=1}^2 \sum _{j=1}^2 \zeta_i^*\zeta_j S_{ik} S^{-1}_{kj}=a_k+ib_k$ as in section~\ref{ssec:diagonalizable}, where $S$ is the matrix that diagonalizes $W$. We obtain:

\begin{align}
\Re(\kappa_1)&=\frac{1}{2} \left(\zeta _1^2+\zeta _2^2+\frac{2 \zeta_1 \zeta_2 (\epsilon +1)}{\sqrt{\epsilon  (\epsilon +2)}}\right)\,,\\[0.2cm]
\Re(\kappa_2)&=\frac{1}{2} \left(\zeta _1^2+\zeta _2^2-\frac{2 \zeta _2 \zeta _1 (\epsilon +1)}{\sqrt{\epsilon  (\epsilon +2)}}\right)\,,
\end{align}
\begin{align}
\Im(\kappa_1)&=\frac{\zeta _1^2-\zeta _2^2}{2 \sqrt{\epsilon  (\epsilon +2)}}\,,\\[0.2cm]
\Im(\kappa_2)&=\frac{\zeta _2^2-\zeta _1^2}{2 \sqrt{\epsilon  (\epsilon +2)}}\,.
\end{align}

Note that each of these diverges as $\epsilon\to 0$.
However, if we construct the spectral density from~\eqref{eqn:Jeff_diagonalizable}, we obtain:
\begin{equation}
    J(\omega)=\frac{4 \delta  \left(\delta  \zeta _1 (\epsilon +1)+\zeta _2 \omega \right){}^2}{\delta ^4 (\epsilon +1)^4-2 \delta ^2 \omega ^2 (\epsilon  (\epsilon +2)-1)+\omega ^4}\,.
\end{equation}
Taking the limit of $\epsilon\to 0$, we obtain:
\begin{equation}
    J(\omega)=\frac{4 \delta  \left(\delta  \zeta _1+\zeta _2 \omega \right){}^2}{\left(\delta ^2+\omega ^2\right)^2}\,,
\end{equation}
exactly as before.

\section{$\mathbf{3\times3}$ non-diagonalizable Configuration\label{apx:ND3}}
In general, constructing an $n\times n$ non-diagonalizable matrix that is the sum of an anti-Hermitian matrix a and real diagonal matrix, as $W$ is, is a nontrivial task.
However, we can use the $2\times2$ non-diagonalizable block to construct a $3\times3$ non-diagonalizable block for 3 pseudomodes arranged in a chain.
We begin with the $2\times2$ non-diagonalizable block and expand it to $3\times3$, leaving the two additional off-diagonal elements as unknown:
\begin{equation}
    W = (-i\varepsilon-\eta)\mathds{1}+\delta\begin{bmatrix}
    0 & -ix & 0\\
    -ix & 0 & -i\\
0 & -i & -2
    \end{bmatrix}\,.
\end{equation}

We compute the eigenvalues of this matrix as:
\begin{equation}\begin{aligned}
\lambda_1(x) &= -\frac{2}{3} - \frac{-1 + 3x^2}{3\left(1 - 18x^2 + 3\sqrt{3} \sqrt{-x^2 + 11x^4 + x^6}\right)^{1/3}}\\[0.2cm] +& \frac{1}{3}\left(1 - 18x^2 + 3\sqrt{3} \sqrt{-x^2 + 11x^4 + x^6}\right)^{1/3} \,,
\end{aligned}\end{equation}

\begin{equation}\begin{aligned}
\lambda_2(x) &= -\frac{2}{3} + \frac{(1 + i\sqrt{3})(-1 + 3x^2)}{6\left(1 - 18x^2 + 3\sqrt{3} \sqrt{-x^2 + 11x^4 + x^6}\right)^{1/3}}\\[0.2cm] -& \frac{1}{6}(1 - i\sqrt{3})\left(1 - 18x^2 + 3\sqrt{3} \sqrt{-x^2 + 11x^4 + x^6}\right)^{1/3} \,,
\end{aligned}\end{equation}

\begin{equation}\begin{aligned}
\lambda_3(x) &= -\frac{2}{3} + \frac{(1 - i\sqrt{3})(-1 + 3x^2)}{6\left(1 - 18x^2 + 3\sqrt{3} \sqrt{-x^2 + 11x^4 + x^6}\right)^{1/3}}\\[0.2cm] -& \frac{1}{6}(1 + i\sqrt{3})\left(1 - 18x^2 + 3\sqrt{3} \sqrt{-x^2 + 11x^4 + x^6}\right)^{1/3}\,.
\end{aligned}\end{equation}\vskip8pt

Plotting the eigenvalues, one can see a point where $\lambda_1(x)$ and $\lambda_3(x)$ coincide and that it occurs where the derivative of both eigenvalues with respect to $x$ is undefined. 
Informed by this plot, we are able to find $x$ where both eigenvalues coincide, and can then verify that the resulting $W$ is non-diagonalizable. 
In particular, we find:
\begin{equation}
    x = \sqrt{\frac{1}{2}\left(5\sqrt{5}-11\right)}\,,
\end{equation}
and so we find a candidate $3\times3$ non-diagonalizable $W$ as:
\begin{equation}\label{eqn:apxe_W}
    W = (-i\varepsilon-\eta)\mathds{1}+\delta\begin{bmatrix}
    0 & \sqrt{\frac{1}{2}\left(5\sqrt{5}-11\right)} & 0\\
    \sqrt{\frac{1}{2}\left(5\sqrt{5}-11\right)} & 0 & -i\\
0 & -i & -2
    \end{bmatrix}\,.
\end{equation}

If we then proceed as in the $2\times2$ case and allow only the first pseudomode to couple to the system and the last pseudomode to couple to the residual bath (so $\eta=0$), we can then compute the resulting effective memory kernel
\begin{equation}\begin{aligned}
    \chi_{\rm eff}(t)&=|\zeta|^2e^{-i\varepsilon t}\left\{\frac{e^{-\left(\sqrt{5}-1\right) \delta  t} \left[\left(7 \sqrt{5}-17\right) e^{\frac{1}{2} \left(3 \sqrt{5}-5\right) \delta  t}+8 \sqrt{5}-18\right]}{5 \left(3 \sqrt{5}-7\right)}\right. \\&\qquad\left.+\frac{\delta  t e^{\frac{1}{2} \left(\sqrt{5}-3\right) \delta  t}}{\sqrt{5}}\right\}\,, \end{aligned}\end{equation}
and effective spectral density\begin{equation}
    J_{\rm eff}(\omega)=\frac{8 (-11 + 5 \sqrt{5}) \delta^5 \lambda_0^2}{\left[ (7-3 \sqrt{5}) \delta^2 + 2 (\omega-\varepsilon)^2 \right]^2 \left[ 2 (3- \sqrt{5}) \delta^2 + (\omega-\varepsilon)^2 \right]}\,,
\end{equation}
for this pseudomode configuration.

We note that there are only terms like $e^t$ and $te^t$ in the effective memory kernel, since $W$ as given by~\eqref{eqn:apxe_W} has two eigenvalues and eigenvectors, rather than just one, and the defective eigenvalue has an algebraic multiplicity of 2.

\section{Inversion Problem for 2 Pseudomodes\label{apx:2x2inversion}}

In this appendix we explore the simple case of two pseudomodes. \edit{While a solution for the pseudomode parameters in this case has been previously shown~\cite{pleasance2020a}, we are able to show using the methods of Sec.~\ref{sec:fitting} that, for the 2-mode case, it is always possible to choose pseudomode parameters such that $\Gamma>0$ for effective spectral density that is nonnegative.}

Suppose we obtain a 2-mode fit to the memory kernel with eigenvalues $\varepsilon_1,\varepsilon_2,\gamma_1,\gamma_2$ and amplitudes $\kappa_1,\kappa_2$. First note that to satisfy~\eqref{eqn:kappa_sum}, the $\kappa$'s must be of the following form:
\begin{equation}
    \kappa_1=\alpha_1+i\beta\,,\quad\kappa_2=\alpha_2-i\beta\,,
\end{equation}
where $\alpha_1,\alpha_2,\beta$ are real numbers and $\alpha_1+\alpha_2>0$. In~\eqref{eqn:vfromu}, we fix
\begin{equation}
    u_j=1\,,\quad v_j=\kappa_j^*
\end{equation}

Since the matrices here are only $2\times2$, we can then construct $S^\dag S$ directly from~\eqref{eqn:vuPlusBperp} as:
\begin{align}
    S^\dag S&=\begin{bmatrix}
        \kappa_1^*/2&\kappa_1^*/2\\
        \kappa_2^*/2&\kappa_2^*/2
    \end{bmatrix} +\begin{bmatrix}
        a&-a\\b&-b
    \end{bmatrix}\,.
\end{align}

Enforcing positivity then requires:
\begin{align}
    (S^\dag S)_{11}>0&\implies a=\frac{\kappa_1}{2}+a'\,,\,a'>-\alpha_1\\
    (S^\dag S)_{22}>0&\implies b=-\frac{\kappa_2}{2}+b'\,,\,b'>-\alpha_2\,.
\end{align}
and then for Hermiticity we must have $b'=a'$. We thus obtain $S^\dag S$ in terms of one free parameter $a'>\max\{0,-\alpha_1,-\alpha_2\}$ as:
\begin{equation}
    S^\dag S=\begin{bmatrix}
        \alpha_1+a' & i\beta+a'\\
        -i\beta-a'&\alpha_2+a'
    \end{bmatrix}\,.
\end{equation}

The final constraint to enforce $S^\dag S>0$ is that we must have
\begin{equation}\label{eqn:positive-det}
    \det(S^\dag S)=a'(\alpha_1+\alpha_2)+\alpha_1\alpha_2-\beta^2>0\,,
\end{equation}
and since we must have $\alpha_1+\alpha_2>0$ from~\eqref{eqn:kappa_sum} we can always choose $a'$ large enough that~\eqref{eqn:positive-det} is satisfied. Since $S^\dag S$ is $2\times2$ we can express $S=\sqrt{S^\dag S}$ analytically as~\cite{levinger1980}:
\begin{equation}
\begin{aligned}
    S&=\frac{1}{\sqrt{2 a+\alpha _1+\alpha _2+2 \sqrt{\det(S^\dag S) }}}\quad\times\\&\begin{bmatrix}
a'+\alpha_1+\sqrt{\det(S^\dag S)}&-a'-i\beta\\
-a'+i\beta & a'+\alpha_2+\sqrt{\det(S^\dag S)}
\end{bmatrix}\,.
\end{aligned}
\end{equation}

Using~\eqref{eqn:W_tilde} we can find $\tilde{\Gamma}$ and compute its eigenvalues, which will be the same as the eigenvalues of the diagonal $\Gamma$ obtained in~\eqref{eqn:inversion_solution}. Those eigenvalues have the form:
\begin{equation}\label{eqn:2modeinversion-eigs}
\frac{\gamma_1+\gamma_2}{2}\pm\frac{\sqrt{
c_1(\gamma_1-\gamma_2)^2+c_2(\varepsilon_1-\varepsilon_2)^2
}}{c_3}\,,
\end{equation}
where $c_1$, $c_2$, $c_3$ are all positive and depend only on $\alpha_1$, $\alpha_2$, $\beta$, and $a'$. From this equation we can see that any choice of $a'$ that makes one of the rates smaller will make the other larger. We can also see that in order to have both rates be positive, the second term must be smaller than the first. Thus, given some $\alpha_1,\alpha_2,\beta$ from a 2-mode Prony's method fit we can immediately tell whether it will be possible for that fit to have both rates be positive by looking at~\eqref{eqn:2modeinversion-eigs} and asking whether there is a choice of $a'$ such that the second term is smaller than the first.

We can simplify further in the case that we have a symmetric spectral density, such that $J(\omega)=J(-\omega)$. In that case we expect a fit from Prony's method to follow the same symmetry, i.e,
\begin{equation}\label{eqn:symmetry-condition}
    \gamma_1=\gamma_2\,,\quad \varepsilon_1=-\varepsilon_2\,,\quad\alpha_1=\alpha_2\,.
\end{equation}

In this case,~\eqref{eqn:2modeinversion-eigs} simplifies greatly to (removing the redundant subscripts)
\begin{equation}
    \gamma\pm\frac{2\sqrt{a'^2+\beta^2}|\varepsilon|}{\sqrt{2a'\alpha+\alpha^2-\beta^2}}\,.
\end{equation}

To see under what conditions we can choose $a'$ such that both rates are positive, we can then simply solve
\begin{equation}\label{eqn:r-pos-eig}
    \frac{2 \sqrt{a'^2+\beta ^2} \left| \varepsilon\right| }{\sqrt{2 a' \alpha+\alpha^2-\beta ^2}}=r\gamma_1\,,
\end{equation}
for any $0\leq r\leq 1$ and see under what conditions we find a real solution for $a'$. Choosing $r=1$ makes one of the rates zero and maximizes the other, while $r=0$ would make both rates identically $\gamma_1$. Squaring both sides of~\eqref{eqn:r-pos-eig}, we can rearrange to obtain an equation that is quadratic in $a'$; the solutions are then
\begin{equation}
    a'=\frac{\alpha r^2\gamma _1^2\pm\sqrt{\left(r^2\gamma^2+4 \varepsilon^2\right) \left(\alpha^2 r^2\gamma^2-4 \beta ^2 \varepsilon^2\right)}}{4 \varepsilon^2}\,,
\end{equation}
and there is only a real solution if
\begin{equation}
    r^2\alpha^2\gamma^2>4 \beta ^2 \varepsilon^2\,.
\end{equation}

This condition is least restrictive when $r=1$ and so we can see that there is only a choice of $a'$ that will result in a $\Gamma$ matrix that is positive if the fit parameters satisfy
\begin{equation}\label{eqn:a'-bound}
    \frac{\alpha^2}{\beta^2}>4\frac{\varepsilon^2}{\gamma^2}\,.
\end{equation}

We now compare this bound with the bounds on fit parameters such that the resulting effective spectral density is positive everywhere. The effective spectral density~\eqref{eqn:Jeff_diagonalizable} corresponding to a fit satisfying~\eqref{eqn:symmetry-condition} is
\begin{equation}
    J_{\rm eff}(\omega)=\frac{\alpha\gamma+2\beta(\omega-\varepsilon)}{(\omega-\varepsilon)^2+(\gamma/2)^2}+\frac{\alpha\gamma-2\beta(\omega+\varepsilon)}{(\omega+\varepsilon)^2+(\gamma/2)^2}\,.
\end{equation}

We assume without loss of generality that $\varepsilon>0$; we then might have $\beta$ to be positive or negative. Computing $dJ_{\rm eff}(\omega)/d\omega=0$, we find up to five potential critical points. One of them is at $\omega=0$, and the others are:
\begin{equation}
\begin{aligned}
    &\omega_{c,\pm,\pm'}=\pm\frac{1}{2}\,\times\\&\sqrt{\frac{\pm'4 \gamma  | \varepsilon |  \sqrt{\left(\alpha ^2+\beta ^2\right) \left(\gamma ^2+4 \varepsilon ^2\right)}-\left(\gamma ^2+4 \varepsilon ^2\right) (\alpha  \gamma -2\beta  \varepsilon )}{\alpha  \gamma +2\beta  \varepsilon }}\,.
\end{aligned}
\end{equation}

Starting with the critical point at zero, we see
\begin{equation}
    J_{\rm eff}(0)=\frac{2\alpha\gamma-4\beta\varepsilon}{\varepsilon^2+\gamma^2/4}\,,
\end{equation}
and so for this to be positive we require
\begin{equation}\label{eqn:ag>be}
    \alpha\gamma>2\beta\varepsilon\,.
\end{equation}

Which of the other possible critical points are real will depend on the sign of $\alpha\gamma+2\beta\varepsilon$. We first consider the case that it is positive, in which case there can only possibly be critical points for $\pm'=+$. We then see that
\begin{equation}
    J_{\rm eff}(\omega_{c,+,+})=\frac{2\alpha\varepsilon-\beta\gamma+\sqrt{(\alpha^2+\beta^2)(\gamma^2+4\varepsilon^2)}}{\gamma\varepsilon}\,,
\end{equation}
which is always positive, since $\sqrt{(\alpha^2+\beta^2)(\gamma^2+4\varepsilon^2)}>|\beta|\gamma$. Thus, $J_{\rm eff}$ will always be positive at these points (if they are real).

Finally, we consider the case that $\alpha\gamma+2\beta\varepsilon<0$, in which case we know $\beta<0$ and there will be at least two critical points where $\pm'=-$. We then obtain
\begin{equation}
    J_{\rm eff}(\omega_{c,+,-})=\frac{2\alpha\varepsilon-\beta\gamma-\sqrt{(\alpha^2+\beta^2)(\gamma^2+4\varepsilon^2)}}{\gamma\varepsilon}\,,
\end{equation}
and for this to be positive, we have:
\begin{align}
2\alpha\varepsilon+|\beta|\gamma&>\sqrt{(\alpha^2+\beta^2)(\gamma^2+4\varepsilon^2)}\,.
\end{align}
Squaring both sides and rearranging, we find we must have
\begin{align}
    (\alpha\gamma-2\beta\varepsilon)^2<0\,,
\end{align}
which is impossible. So we have the additional constraint that we must have $\alpha\gamma+2\beta\varepsilon>0$. Combining this with~\eqref{eqn:ag>be}, we see that we can write
\begin{equation}
    \alpha\gamma>2|\beta|\varepsilon\,,
\end{equation}
or, to mirror~\eqref{eqn:a'-bound}:
\begin{equation}\label{eqn:pos-bound}
    \frac{\alpha^2}{\beta^2}>4\frac{\varepsilon^2}{\gamma^2}\,.
\end{equation}

This exactly matches the bound~\eqref{eqn:a'-bound}! This means that, given a set of fit parameters $\alpha,\beta,\gamma,\varepsilon$, we can find an $a'$ such that $\Gamma>0$ if and only if the effective spectral density from those fit parameters is positive.

\section{Fit Parameters in Figures~\ref{fig:comparefits} \edit{and~\ref{fig:numex}\label{apx:data}}}
All numerical data in this article are shown, or can be reproduced from, the data contained in the following tables.
\begin{widetext}

\begin{table}[!h]
\caption{Fit parameters for the diagonal fit used in Figs.~\ref{fig:comparefits}(a) and~\ref{fig:comparefits}(b), as they appear in~\eqref{eqn:Jeff_diagonal_simple}.}
\label{tab:data1}
\begin{ruledtabular}
\begin{tabular}{ccc}
$\kappa_k$ & $\varepsilon_k$ & $\gamma_k$\\\hline
0.04934 & 0.47611 & 0.33539 \\
0.04934 & -0.47611 & 0.33539 \\
0.06613 & -0.16441 & 0.39853 \\
0.06613 & 0.16441 & 0.39853 \\
0.02641 & 0.7457 & 0.23523 \\
0.02641 & -0.7457 & 0.23523
\end{tabular}
\end{ruledtabular}
\end{table} %

\begin{table}[!h]
\caption{Fit parameters for the diagonalizable fit used in Figs.~\ref{fig:comparefits}(a) and~\ref{fig:comparefits}(c), as they appear in~\eqref{eqn:Jeff_diagonalizable}.}
\label{tab:data2}
\begin{ruledtabular}
\begin{tabular}{ccc}
$\kappa_k$ & $\varepsilon_k$ & $\gamma_k$\\\hline
0.10807 + 0.05924$i$ & 0.34457 & 1.23329 \\
0.10807 - 0.05924$i$ & -0.34457 & 1.23329 \\
0.01708 - 0.05127$i$ & -0.79313 & 0.62406 \\
0.01708 + 0.05127$i$ & 0.79313 & 0.62406 \\
0.00014 - 0.01174$i$ & -0.95949 & 0.15409 \\
0.00014 + 0.01174$i$ & 0.95949 & 0.15409
\end{tabular}
\end{ruledtabular}
\end{table}

\begin{table}[!h]
\caption{\edit{Fit parameters for the uncoupled modes fit used in Fig.~\ref{fig:numex}, as they appear in~\eqref{eqn:Jeff_diagonal_simple}.}}
\label{tab:data3}
\begin{ruledtabular}
\begin{tabular}{ccc}
$\kappa_k$ & $\varepsilon_k$ & $\gamma_k$\\\hline
0.0369 & -0.706 & 0.295 \\
0.0369 &  0.706 & 0.295 \\
0.0699 & -0.377 & 0.423 \\
0.0699 &  0.377 & 0.423 \\
0.0860 & 1.1$\times10^{-10}$ & 0.476 \\
3.9$\times10^{-11}$ & 4.753 & 0.0819
\end{tabular}
\end{ruledtabular}
\end{table}

\begin{table*}[!h]
\caption{\edit{Elements $\Lambda_{ij}$ for the pseudomode configuration shown in Fig.~\ref{fig:numex}, as they appear in~\eqref{eqn:Jeff}.}}
\label{tab:data-numex}
\begin{ruledtabular}
\begin{tabular}{c|c|cccccc}
$\Gamma$ & $\zeta$ & & & $\Lambda_{ij}$ & & & \\\hline
-0.086  & -0.222 - 0.318$i$ & 0.0738 & 0.755-0.00389$i$ & -0.131-0.00188$i$ & 0.0696-0.122$i$ & 0.164+0.237$i$ & 0.0427-0.00337$i$ \\
7.6$\times10^{-5}$ & 0.119 - 0.0121$i$ & 0.755+0.00389$i$ & 0.00300 & 0.210-0.0581$i$ & -0.227-0.258$i$ & 0.239-0.250$i$ & 0.131-0.253$i$ \\
6.8$\times10^{-6}$ & 0.0710 + 0.200$i$ & -0.131+0.00188$i$ & 0.210+0.0581$i$ & -0.0894 & 0.395-0.404$i$ & -0.255-0.434$i$ & 0.337+0.174$i$ \\
0.962 & -0.136 - 0.0644$i$ & 0.0696+0.122$i$ & -0.227+0.258$i$ & 0.395+0.404$i$ & 0.000761 & 0.0167-0.000388$i$ & 0.148-0.0934$i$ \\
1.383 & 0.0132 + 0.169$i$ & 0.164-0.237$i$ & 0.239+0.250$i$ & -0.255+0.434$i$ & 0.0167+0.000388$i$ & 0.0140 & -0.481-0.485$i$ \\
6.575 & -0.103 + 0.117$i$ & 0.0427+0.00337$i$ & 0.131+0.253$i$ & 0.337-0.174$i$ & 0.148+0.0934$i$ & -0.481+0.485$i$ & -0.00211
\end{tabular}
\end{ruledtabular}
\end{table*}
\color{white}
\vspace{-4ex}
\end{widetext}
\color{black}$\,$
\bibliography{references}


\end{document}